\documentclass{elsarticle}

\usepackage[reviewcopy]{adndt}
\usepackage{longtable}

\usepackage{amsmath}

\usepackage{amssymb}

\biboptions{square,sort&compress}
\bibpunct[]{[}{]}{,}{n}{}{;}
\begin{document}

\begin{frontmatter}

\journal{Atomic Data and Nuclear Data Tables}


\title{Discovery of the Isotopes with Z $\le$ 10}

  \author{M. Thoennessen\corref{cor1}}\ead{thoennessen@nscl.msu.edu}

  \cortext[cor1]{Corresponding author.}

  \address{National Superconducting Cyclotron Laboratory and \\ Department of Physics and Astronomy, Michigan State University, \\ East Lansing, MI 48824, USA}

\date{April 10, 2010} 

\begin{abstract}
A total of 126 isotopes with Z $\le$ 10 have been identified to date. The discovery of these isotopes which includes the observation of unbound nuclei, is discussed. For each isotope a brief summary of the first refereed publication, including the production and identification method, is presented.
\end{abstract}

\end{frontmatter}





\newpage
\tableofcontents
\listofDfigures
\listofDtables

\vskip5pc

\section{Introduction}\label{s:intro}

The discovery of the light isotopes is discussed as part of the series of the discovery of isotopes which began with the cerium isotopes in 2009 \cite{2009Gin01}. The purpose of this series is to document and summarize the discovery of all isotopes. Guidelines for assigning credit for discovery are (1) clear identification, either through decay-curves and relationships to other known isotopes, particle or $\gamma$-ray spectra, or unique mass and Z-identification, and (2) publication of the discovery in a refereed journal. The authors and year of the first publication, the laboratory where the isotopes were produced as well as the production and identification methods are discussed. When appropriate, references to conference proceedings, internal reports, and theses are included. When a discovery includes a half-life measurement the measured value is compared to the currently adopted value taken from the NUBASE evaluation \cite{2003Aud01} which is based on the ENSDF database \cite{2008ENS01}.

\section{Discovery of Isotopes with Z $\le$ 10}

The discovery of 125 isotopes with Z $\le$ 10 includes 2 neutral, 7 hydrogen, 9 helium, 10 lithium, 9 beryllium, 13 boron, 14 carbon, 14 nitrogen, 14 oxygen, 16 fluorine, and 18 neon isotopes. Most likely, all nuclei which are stable with respect to neutron or proton emission have been observed. The only isotopes that could potentially be stable are $^{33}$F and $^{36}$Ne. Thus the neutron drip-line has been reached up to Z = 8.

The discovery of the light stable isotopes is not easily defined because they were involved in the discovery of isotopes themselves. We decided to credit discovery if the detection method was sensitive enough to separate isotopes and if isotopes were specifically searched for. Thus the first description of a mass spectrograph by Dempster in 1918 was not considered because Dempster observed only a single isotope per element and did not perform absolute mass measurements \cite{1918Dem01}.

\begin{figure}
	\centering
	\includegraphics[scale=0.9]{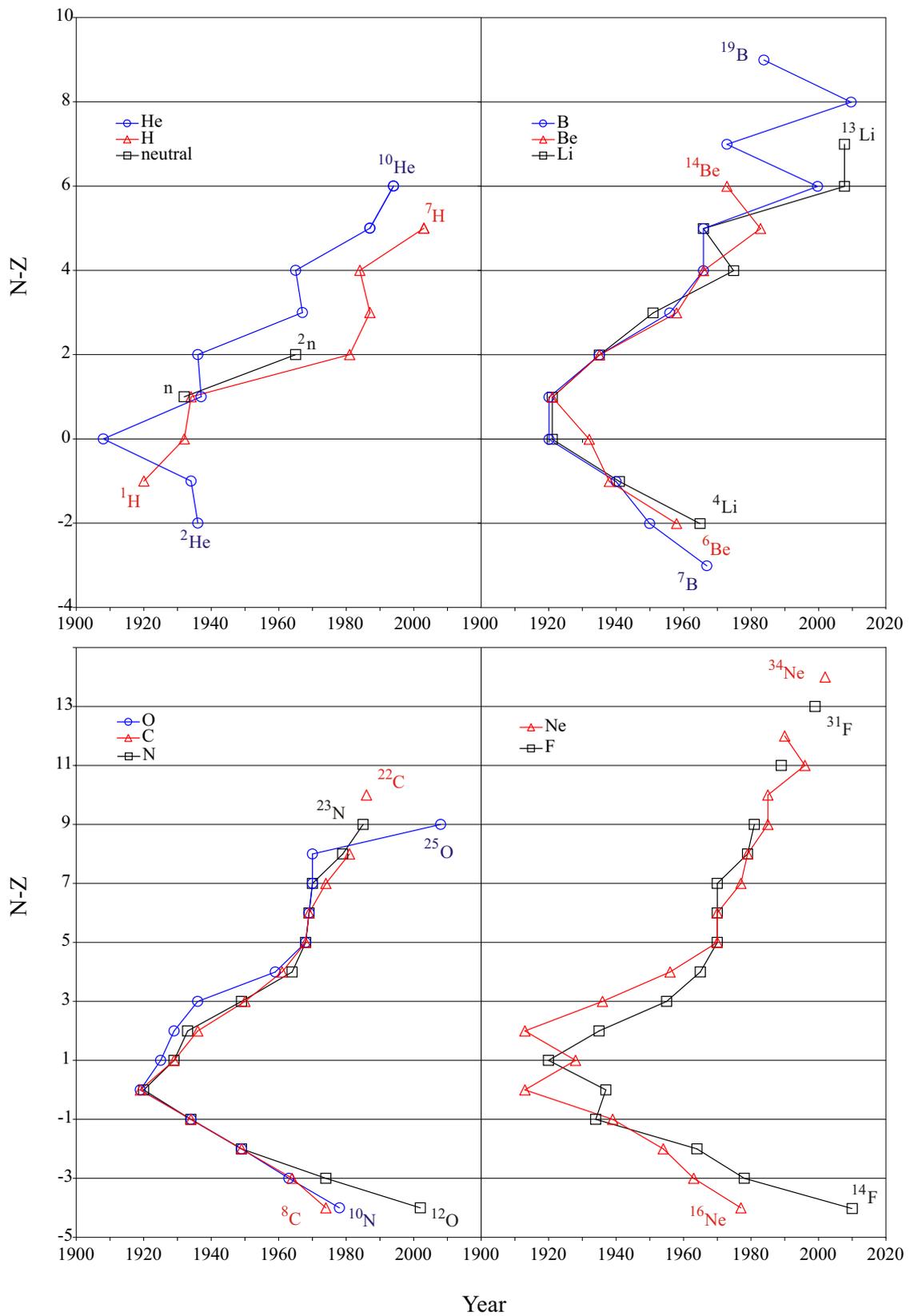}
	\caption{Isotopes as a function of year when they were discovered. The vertical axis is shown as N$-$Z.}
\label{f:year}
\end{figure}

Nuclei beyond the driplines, i.e. nuclei which decay by the emission of a neutron or a proton were included in the compilation. In some cases these nuclei live for only very short times and especially for nuclei beyond the neutron dripline they can only be measured as resonances. Nevertheless the masses can be determined by transfer reactions or by invariant mass measurements and the lifetimes can be determined from the width of the resonance. However, especially for nuclei which are removed by two or more neutrons from the last particle-stable isotope, these resonant states can be very broad and it becomes questionable if the corresponding lifetimes are long enough to be called a nucleus \cite{2004Tho01}. Only unbound nuclei for which a resonance was observed are included, the first ``non-existence'' for an unbound nucleus has been compiled elsewhere \cite{2004Tho01}.

For heavier isotopes we have adopted the practice to accept the observation of isomeric states or excited proton unbound states prior to the determination of the ground states as the discovery. Accordingly for the light nuclei the observed resonance does not necessarily have to correspond to the ground state.

Figure \ref{f:year} summarizes the year of first discovery for all light isotopes. The isotopes are shown on the vertical axis as (N-Z). The radioactive isotopes were produced using light-particle reactions (LP), heavy-ion transfer reactions (TR), fusion-evaporation reactions (FE), deep-inelastic reactions (DI), pion-induced reactions (PI), spallation (SP), projectile fragmentation of fission (PF), and most recently with secondary beams (SB). The stable isotopes were identified using cathode rays (CR), atomic (AS) and mass spectroscopy (MS), and light-particle reactions (LP). Heavy ions are all nuclei with an atomic mass larger than A=4 \cite{1977Gru01}. Light particles also include neutrons produced by accelerators. In the following, the discovery of each isotope is discussed in detail. Specifics of the discoveries are listed in Table 1.

\subsection{Z = 0}\vspace{0.0cm}

In addition to the discovery of the neutron, we also included the first measuring of the s-wave scattering length of the di-neutron system as a ``virtual'' resonance in the list. The tentative indication of a bound tetra-neutron \cite{2002Mar01} was not included because it was not confirmed in subsequent experiments \cite{2005Ale01}. In addition, parts of the analysis were questioned \cite{2004She01} and the existence of a bound tetra-neutron seems theoretically very unlikely \cite{2003Pie01}.

\subsubsection*{$^1$n}

The neutron was discovered by Chadwick in 1932 as described in the paper ``Possible Existence of a Neutron'' \cite{1932Cha01}. Beryllium was bombarded with $\alpha$-particles and the resulting radiation was measured with an ionization chamber. ``[The] results, and others I have obtained in the course of the work, are very difficult to explain on the assumption that the radiation from beryllium is a quantum radiation, if energy and momentum are to be conserved in the collisions. The difficulties disappear, however, if it be assumed that the radiation consists of particles of mass 1 and charge 0, or neutrons. The capture of the $\alpha$-particle by the Be$^9$ nucleus may be supposed to result in the formation of a C$^{12}$ nucleus and the emission of the neutron.''

\subsubsection*{$^2$n}

The first accurate measurement of the unbound two-neutron resonance state was reported in 1965 by Haddock et al. in ``Measurement of the Neutron-Neutron s-Wave Scattering Length from the Reaction $\pi^- + d \rightarrow 2n + \gamma$'' \cite{1965Had01}. Negative pions produced by the Berkeley 184-inch cyclotron bombarded a liquid deuterium target and the reaction $\pi^- + d \rightarrow 2n + \gamma$ was studied. Neutrons and $\gamma$-rays were detected by scintillation detectors. ``The experimental uncertainties in timing resolution, counter efficiency, and background are combined with the result of the $\chi^2$ fit to yield a preliminary value of a$_{nn}$ = $-$16.4$\pm$1.9~F.'' We credit Haddock et al. with the first observation because it represents the first accurate experimentally complete experiment as stated by von Witsch et al.: ``However, for a long time the result of just one such experiment could be considered experimentally accurate as well as reliable from a theoretical point of view, namely the kinematically complete measurement $2$H($\pi^-$,2n)$\gamma$ by Haddock et al. ... In all other reactions at least one additional nucleon was present in the final state whose interaction with the two neutrons could not be treated in a dynamically exact way'' \cite{1979Wit01}.

\subsection{Hydrogen}\vspace{0.0cm}

In addition to the two stable hydrogen isotopes $-$ the proton and the deuteron $-$ and the radioactive tritium, four additional neutron-rich unbound resonances up to $^7$H have been reported. The assignments for the discoveries of these resonances is certainly arguable because they are broad and sometimes different techniques observed different resonances. A consistent description is not available for these resonances at the present time and the assignments should be considered as tentative.

\subsubsection*{$^1$H}

In 1920 $^1$H was first measured in a mass spectrometer by Aston in ``The Constitution of the Elements'' \cite{1920Ast01}. ``By the same methods H$_3$, H$_2$, and H$_1$ all give consistent results for the mass of the hydrogen atom as 1.008 within experimental error, agreeing with the value given by chemical analysis...''.

\subsubsection*{$^2$H}

Urey discovered $^2$H as described in the publication ``A Hydrogen Isotope of Mass 2'' in 1932 \cite{1932Ure01}. Liquid hydrogen was evaporated near its triple point in order to enhance the concentration of heavy hydrogen. The atomic spectra were then observed in a hydrogen discharge tube. ``Under these conditions we found in this sample as well as in ordinary hydrogen faint lines at the calculated positions for the lines of H$^2$ accompanying H$_\beta$, H$_\gamma$, H$_\delta$. These lines do not agree in wave-length with any molecular lines reported in the literature.''

\subsubsection*{$^3$H}

In 1934 Oliphant et al. reported the discovery of $^3$H ``Transmutation Effects observed with Heavy Hydrogen'' \cite{1934Oli01}. Deuterated ammonium chloride, ammonium sulphate, and orthophosphoric acid samples were bombarded with deuterons accelerated by 20 kV at the Cavendish Laboratory in Cambridge, UK and the range of observed protons were measured. ``It seems more probable that the diplons unite to form a new helium nucleus of mass 4.0272 and 2 charges. This nucleus apparently finds it difficult to get rid of its large surplus energy above that of an ordinary He nucleus of mass 4.0022, but breaks up into two components. One possibility is that it breaks up according to the reaction D$^2_1$ + D$^2_1  \to$ H$^3_1$ + H$^1_1$. The proton in this case has the range of 14 cm. while the range of 1.6 cm. observed agrees well with that to be expected from momentum relations for an H$^3$ particle. The mass of this new hydrogen isotope calculated from mass and energy changes is 3.0151.''

\subsubsection*{$^4$H}

In 1981 Sennhauser described the observation of an unbound resonance of $^4$H in ``Observation of Particle Unstable $^4$H in Pion Absorption in $^7$Li'' \cite{1981Sen01}. Negative pions produced at the Swiss Institute for Nuclear Research in Villigen bombarded a natural lithium target to form $^4$H in the reaction $\pi^- + ^7$Li $\to ^3$H + $^4$H, which subsequently decays into $^3$H and a neutron. The tritons were identified with two collinear solid state $\Delta$E$-$E telescopes. ``The histogram shows the experimental data, the dashed line the three-body phase space contribution (18\%) and the solid line the fit of the calculated spectra to the experimental data, determined by an $^4$H resonance at (2.7$\pm$0.6) MeV with a reduced width $\gamma^2$ = (2.3$\pm$0.6)MeV.''

\subsubsection*{$^5$H}

An unbound resonance in $^5$H was first observed by Gornov et al. in ``Detection of superheavy hydrogen isotopes in the reaction for the absorption of $\pi^-$ mesons by $^9$Be nuclei'' in 1987 \cite{1987Gor01}. Negative pions from the Leningrad Institute of Nuclear Physics synchrocyclotron were stopped in a beryllium target. Charged particles were detected and identified in a multibeam semiconductor spectrometer. ``The experimental distribution can be reproduced satisfactorily by introducing a three-body channel with the formation of a resonant $^5$H state. The parameters of the resonant state of $^5$H in this case are E$_r$ = 7.4$\pm$0.7 MeV and $\Gamma$ = 8$\pm$3 MeV...'' A previous report that $^5$H was bound with respect to neutron emission \cite{1963Nef02} could not be confirmed \cite{1962Cen01,1964Sch01,1964She01}.

\subsubsection*{$^6$H}

The first observation of $^6$H was reported in 1984 by Aleksandrov et al. in ``Observation of an unstable superheavy hydrogen isotope $^6$H in the reaction $^7$Li($^7$Li,$^8$B)'' \cite{1984Ale01}. $^7$Li ions were accelerated to 82 MeV by the Kurchatov Institute isochronous cyclotron and bombarded an enriched $^7$Li target. Charged particles were detected and identified with semiconductor detectors by measuring energy-loss, energy and time of flight. ``Thus, the maximum observed at an energy 35$-$40 MeV of the $^8$B nuclei is due to formation of the unstable nucleus $^6$H... The mass defect is 41.9$\pm$0.4 MeV, from which it follows that $^6$H is unstable against the decay $^6$H $\to$ $^3$H + 3n by 2.7$\pm$0.4 MeV, and the width is $\Gamma$ = 1.8$\pm$0.5 MeV, which gives for the $^6$H lifetime a value 3.7$\cdot$10$^{-22}$ sec.''

\subsubsection*{$^7$H}

In 2003 Korsheninnikov et al. observed $^7$H in ``Experimental Evidence for the Existence of $^7$H and for a Specific Structure of $^8$He'' \cite{2003Kor01}. A cryogenic hydrogen target was bombarded with a 61.3~MeV/u secondary $^8$He beam produced with the fragment separator RIPS at RIKEN, Japan. $^7$H was produced in the reaction p($^8$He,pp)$^7$H and identified by measuring the two protons in coincidence. ``As seen in [the figure], the experimental spectrum increases near the t + 4n threshold even more sharply than the most extreme curve. This provides a strong indication on the possible existence of $^7$H state near the t + 4n threshold.''

\subsection{Helium}\vspace{0.0cm}

Nine helium isotopes are described including 2 stable, 2 neutron-rich, 4 neutron-unbound resonances as well as the di-proton. $^4$He is certainly a special case where we assigned Rutherford's determination of the mass and charge of the $\alpha$-particle as the discovery.

\subsubsection*{$^2$He}

The 1936 paper ``The scattering of protons by protons'' by Tuve et al. can be considered as the first evidence of a virtual state in $^{2}$He  \cite{1936Tuv01}. The angular distribution from 15$^\circ$ and 45$^\circ$ of scattered protons were measured for incident proton energies of 600, 700, 800, and 990~keV. ``Measurement of the scattering of protons by deuterium, helium, and air, together with ``vacuum-scattering'' tests which eliminate slit scattering and unknown vapors, have led to the conclusion that the observed anomaly is not due to a contamination and must be ascribed to a proton-proton interaction at close distances (less that 5$\times$10$^{-13}$cm) which involves a marked departure from the ordinary Coulomb forces.'' The theoretical interpretation was presented by Breit et al. in an accompanying paper \cite{1936Bre01}.

\subsubsection*{$^3$He}

In 1934 Oliphant et al. reported the discovery of $^3$He ``Transmutation Effects observed with Heavy Hydrogen'' \cite{1934Oli01}. Deuterated ammonium chloride, ammonium sulphate, and orthophosphoric acid samples were bombarded with deuterons accelerated by 20 kV at the Cavendish Laboratory in Cambridge, UK, and the range of observed protons were measured. In addition, a large number of neutrons were observed. ``It seems more probable that the diplons unite to form a new helium nucleus of mass 4.0272 and 2 charges. This nucleus apparently finds it difficult to get rid of its large surplus energy above that of an ordinary He nucleus of mass 4.0022, but breaks up into two components... Another possible reaction is D$^2_1$ + D$^2_1  \to$ He$^3_2$ + n$^1_0$ leading to the production of a helium isotope of mass 3 and a neutron. In a previous paper we suggest that a helium isotope of mass 3 is produced as a result of the transmutation of Li$^6$ under proton bombardment into two doubly charged particles. If this last reaction be correct, the mass of He$^3_2$ is 3.0165, and using this mass and Chadwick's mass for the neutron, the energy of the neutron comes out to be about 3 million volts.'' The quoted previous paper only suggests the formation of $^3$He as one possibility \cite{1933Oli01}.

\subsubsection*{$^4$He}

Rutherford's 1908 publication of ``The Charge and Nature of the $\alpha$-Particle'' can be considered as the first identification of the $^4$He isotope \cite{1908Rut01}. An $\alpha$-particle source of ``radium C'' ($^{214}$Bi) was placed inside an electromagnetic field to measure its charge. From the known E/M ratio the mass was deduced. ``We have already seen that the evidence is strongly in favour of the view that E = 2e. Consequently M = 3.84~m, i.e., the atomic weight of an $\alpha$-particle is 3.84. The atomic weight of the helium atom is 3.96. Taking into account probable experimental errors in the estimates of the value of E/M for the $\alpha$-particle, we may conclude that an $\alpha$-particle is a helium atom, or, to be more precise, the $\alpha$-particle, after it has lost its positive charge, is a helium atom.''

\subsubsection*{$^5$He}

The instability of $^5$He was first observed by Williams et al. in ``Evidence for the Instability of He$^5$'' in 1937 \cite{1937Wil01}. $^5$He was produced in the reaction $^7$Li + $^2$H $\to$ $^5$He + $^4$He. Deuterons were accelerated with the 275~kV transformer-kenetron apparatus of the University of Minnesota and the range of $\alpha$-particles was measured with a modified Dunning ionization chamber. ``Assuming the peak of 7.10 cm mean range to be $\alpha$ particles from [$^7$Li + $^2$H $\to$ $^5$He + $^4$He], the expected range of He$^5$ is 4.35 cm on reasonable assumptions for the range-energy distribution for He$^5$ as compared to He$^4$. The mass of He$^5$ is then 5.0140 which is unstable by 0.93 MeV.''

\subsubsection*{$^6$He}

$^6$He was first observed by Bjerge and Brostr\"om in 1936 as reported in the paper ``$\beta$-Ray Spectrum of Radio-Helium'' \cite{1936Bje01}. Neutrons from a beryllium-radon source bombarded a beryllium target and $\beta$-rays were detected with a Wilson chamber located in a magnetic field. ``The most reasonable assumption as to the formation and disintegration of radio-helium are the processes: $^9_4$Be + $^1_0$n $\to$ $^6_4$He + $^4_2$He and $^6_2$He $\to$ $^6_3$Li + e$^-$. If the energy release in [the decay] is 3.7 m.e.v., the mass of $^6_2$He would be 6.0207...'' It is interesting to note that Bjerge already published the correct half-life (0.9(2)~s) earlier, however, he stated ``But it can be said that if the maximum energy is greater than 5.5$\times$10$^6$ e.v., the active body can scarcely be $^6$He, as its mass would then be greater than that of $^4$He plus two neutrons.'' \cite{1936Bje02}.

\subsubsection*{$^7$He}

In 1967 Stokes and Young reported the discovery of $^7$He in ``New Isotope of Helium: $^7$He'' \cite{1967Sto01}. Tritons, accelerated to 22 MeV by the Los Alamos three-stage tandem accelerator bombarded an isotopically pure $^7$Li target. $^7$He was produced in the transfer reaction $^7$Li(t,$^3$He)$^7$He and identified by measuring the reaction product $^3$He in a $\Delta$E gas proportional counter and a surface-barrier E detector. ``The average value of Q obtained by this procedure was $-$11.16 MeV. This Q value corresponds to a mass excess of 26.09$\pm$0.06 MeV ($^{12}$C scale) which has the consequence that $^7$He is unbound to neutron decay by 0.42$\pm$0.06 MeV.''

\subsubsection*{$^8$He}

The first observation of $^8$He was described by Poskanzer et al. in the 1965 paper ``Decay of He$^8$'' \cite{1965Pos01}. Plastic foils or absorbent cotton fibers were irradiated by 2.2 GeV protons from the Brookhaven Cosmotron. The produced $^8$He was transported by a helium carrier gas into a counting chamber where the $\beta$-ray activity was measured with a plastic scintillator. ``A short-lived component was found which decayed to a long-lived tail equal to only one percent of the initial activity. From least-squares fits to all of the decay curves, our best value of the half-life of the short-lived component is 122$\pm$2~msec.'' This half-life is included in the determination of the currently accepted value of 119.1(12)~ms. A previously reported half-life of 30(20)~ms \cite{1963Nef01} could not be confirmed.

\subsubsection*{$^9$He}

Seth et al. reported the first observation of $^9$He in 1987 in ``Exotic Nucleus Helium-9 and its Excited States'' \cite{1987Set01}. Negative pions from the Los Alamos Meson Physics Facility (LAMPF) bombarded a metallic beryllium target and $^9$He was produced in the pion double charge exchange reaction $^9$Be($\pi^-,\pi^+$). A missing mass spectrum was calculated by measuring the positive pions in a magnetic spectrometer. ``...The resulting absolute scale for missing mass leads to the atomic Q$_0$($^9$He(g.s.)) = $-$29.45$\pm$0.10 MeV. This Q value corresponds to an atomic mass excess of 40.80$\pm$0.10 MeV, which implies that the ground state is unstable against single-neutron decay by 1.13$\pm$0.10 MeV.''

\subsubsection*{$^{10}$He}

$^{10}$He was first identified in 1994 by Korsheninnikov et al. in ``Observation of $^{10}$He'' \cite{1994Kor01}. Carbon and CD$_2$ targets were bombarded with a 61~MeV/u secondary $^{11}$Li beam produced with the fragment separator RIPS at RIKEN, Japan. $^{10}$He could be produced in the $^{12}$C($^{11}$Li,$^{10}$He)X or $^2$H($^{11}$Li,$^{10}$He)$^3$He reaction. The excitation energy spectrum of $^{10}$He was calculated from triple coincidence measurements of $^8$He and two neutrons. ``Experimental data show a peak that can be explained as a resonance $^{10}$He, which is unbound by 1.2$\pm$0.3 MeV and its width is less than 1.2 MeV.''

\subsection{Lithium}\vspace{0.0cm}

Ten lithium isotopes are described including 2 stable, 3 neutron-rich, 2 proton- and 3 neutron-unbound resonances. Lithium is the heaviest system for which most likely all isotopes and resonances have been observed.

\subsubsection*{$^4$Li}

An unbound state of $^4$Li was first reported by Cerny et al. in ``Li$^4$ and the excited levels of He$^4$'' \cite{1965Cer01}. The Berkeley 88-in. cyclotron was used to accelerate protons to 43.7 MeV which then bombarded a target of separated $^6$Li. $^4$Li was produced in the (p,t) reaction and identified with a (dE/dx)-E counter telescope. ``[The figure] presents a Li$^6$(p,t)Li$^4$ spectrum at 15$^\circ$. Such data, taken between 10$^\circ$ and 35$^\circ$ in the laboratory, show a broad state which is unbound by 2.9$\pm$0.3 MeV to He$^3$-p decay. (Though we shall denote the peak as the Li$^4$ ground state throughout this report, it is probably not a single state.) The width of the unbound Li$^4$ state is 5.0$\pm$0.5 MeV at all angles.''

\subsubsection*{$^5$Li}

In the 1941 paper ``The Scattering of One- to Three-Mev Protons by Helium'' Heydenburg and Ramsey observed an unbound resonance in $^5$Li \cite{1941Hey01}. Protons from 1.2 to 3.0~MeV accelerated by the Carnegie Institution of Washington pressure electrostatic generator were scattered off a helium gas target and detected with a parallel-plate ionization chamber. ``[From the figure] it appears that the amount of scattering does pass through a maximum at about two Mev as expected theoretically on the basis of the neutron-helium scattering results and the assumption of the equality of n-n and p-p forces. However, the height and narrowness of the maximum are much less than in the neutron-helium case, being only a factor of 2 instead of 5 in intensity, while the half-width is more than one 1 Mev instead of about 0.5 Mev.''

\subsubsection*{$^{6,7}$Li}

In 1921, $^6$Li and $^7$Li were identified for the first time by Aston and Thomson in ``The Constitution of Lithium'' \cite{1921Ast01}. With an externally heated anode at very low pressures it was possible to isolate metallic rays in the mass spectrograph. ``The foregoing results appear to leave no doubt that lithium is a complex element with isotopes of atomic weights 6 and 7.''

\subsubsection*{$^8$Li}

Crane et al. discovered $^8$Li in 1935 as reported in the paper ``The Emission of Negative Electrons from Lithium and Fluorine Bombarded with Deuterons'' \cite{1935Cra02}. Deuterons of 0.8 MeV bombarded a lithium target inside a cloud chamber and $^8$Li was probably formed in the reaction $^7$Li(d,p). Photographs of the electron tracks were taken after the beam was turned off with a switching device. ``To determine the half-life of the active constituent, we adjusted the timing device so that the ion beam was shut off at 1/4, 1/2, 3/4 and 1 second before the chamber expansion. 50 Photographs were taken at each of these settings, and the average numbers of tracks per photograph were found to be 7.08, 4.84, 3.70 and 2.45, respectively. These, plotted on a log scale lie quite closely on a straight line, and indicate a half-life of 0.5$\pm$0.1 second.'' This half-life is close to the currently accepted value of 840.3(9)~ms.

\subsubsection*{$^9$Li}

Gardner et al. described the discovery of $^9$Li in the 1951 paper ``Li$^9$ $-$ New Delayed Neutron Emitter'' \cite{1951Gar01}. Beryllium and boron targets were bombarded with deuterons and protons, respectively, from the Berkeley 184-inch cyclotron. $^9$Li was then formed in the reactions $^9$Be(d,2p) and $^{11}$B(p,3p). The delayed neutron activity after irradiation was measured by photographing the pulses on an oscilloscope. ``The results of these measurements give a half-life ... of T$_{1/2}$ = 0.168$\pm$0.004 sec.'' This value is close to the presently adopted value of 178.3(4)~ms.

\subsubsection*{$^{10}$Li}

$^{10}$Li was first observed by Wilcox et al. in 1975 in ``The ($^9$Be,$^8$B) Reaction and the Unbound Nuclide $^{10}$Li'' \cite{1975Wil01}. A beam of 121 MeV $^9$Be accelerated by the Berkeley 88-inch cyclotron bombarded a $^9$Be target and $^{10}$Li was produced in the transfer reaction ($^9$Be,$^9$B). An unbound state of $^{10}$Li was identified by measuring the energy-loss and energy of the $^8$B ejectiles. ``The observed Q-value for the $^9$Be($^9$Be,$^8$B)$^{10}$Li ground-state reaction was $-$34.06$\pm$0.25 MeV, corresponding to a mass excess for $^{10}$Li of 33.83$\pm$0.25 MeV. The nucleus $^{10}$Li is thus unbound to $^9$Li plus a neutron by 0.80$\pm$0.25 MeV, somewhat more unbound than the current prediction of 0.21 MeV based on the Garvey-Kelson method.''

\subsubsection*{$^{11}$Li}

Poskanzer et al. discovered $^{11}$Li in 1966 in ``New Isotopes: $^{11}$Li, $^{14}$B, and $^{15}$B'' \cite{1966Pos01}. Uranium foils were bombarded with 5.3~GeV protons from the Berkeley Bevatron. Phosphorus-diffused silicon transmission detectors were used in a telescope consisting of an energy-loss, energy, and rejection detector to identify the isotopes. ``The predicted locations of the observed $^{14}$B and $^{15}$B peaks are indicated on the figure by arrows as are the expected positions of the neighboring isotopes $^9$C and $^{10}$C. An additional peak at the predicted location for $^{11}$Li was also observed... the generally reliable calculations of Garvey and Kelson predicted $^{11}$Li to be unbound by 2.5 MeV. Our observation of this isotope was, therefore, very surprising.''

\subsubsection*{$^{12,13}$Li}

Unbound states of $^{12}$Li and $^{13}$Li were observed for the first time by Aksyutina et al. in the 2008 paper ``Lithium isotopes beyond the drip line'' \cite{2008Aks01}. The isotopes were produced in knockout reactions from a $^{14}$Be secondary beam on a liquid hydrogen target. The 304 MeV/u $^{14}$Be beam was produced with the heavy-ion synchroton SIS and selected with the fragment separator FRS. $^{12}$Li and $^{13}$Li were identified with the ALADIN-LAND setup and their relative-energy spectrum was measured. ``This work reports on the first observation of $^{12}$Li and is interpreted as a virtual s-state with a scattering length of $-$13.7(16)~fm... The $^{11}$Li + 2n data give first indication for the existence of a $^{13}$Li resonance at 1.47(31)~MeV.''

\subsection{Beryllium}\vspace{0.0cm}

Nine beryllium isotopes are described including 1 stable, 1 proton-rich, 4 neutron-rich, 1 proton- and 1 neutron-unbound resonance plus $^8$Be which breaks up into two $\alpha$-particles. The one- and two-neutron unbound resonances of $^{15}$Be and $^{16}$Be, respectively, should still be able to be observed.

\subsubsection*{$^6$Be}

An unbound state of $^6$Be was first observed by Bogdanov et al. in 1958 ``The (p,n) reaction on lithium and the ground state of the $^6$Be nucleus'' \cite{1958Bog01}. 9.6~MeV protons from the U.S.S.R. Academy of Sciences Nuclear Energy Institute 1.5~m cyclotron bombarded an enriched $^6$Li target and $^6$Be was produced in the (p,n) charge exchange reaction. $^6$Li was identified by measuring the neutron time-of-flights in stilbene and tolane crystal scintillation counters. ``The Q-value of the reaction $^6$Li(p,n)$^6$Be (ground state) deduced from the experimental data is: Q = $-$5.2$\pm$0.2 MeV and the mass defect of $^6$Be is 20.3$\pm$0.2 MeV.'' A previously reported half-life assigned to either $^6$Be or $^4$Li of 0.4 ~s \cite{1954Tyr01} could not be confirmed.

\subsubsection*{$^7$Be}

Roberts et al. discovered $^7$Be in 1938: ``Radioactivity of $^7$Be'' \cite{1938Rob01}. Deuterons accelerated to 1 MeV bombarded a LiF target and $^7$B was produced in the reaction $^6$Li + $^2$H $\to$ $^7$Be + n. The decay curve was measured following a one-month long irradiation. ``The activity was followed for a month and showed a half-life of 43$\pm$6 days.'' This half-life is close to the presently accepted value of 53.22(6)~d.

\subsubsection*{$^8$Be}

First evidence for $^8$Be was reported by Cockcroft and Walton in the 1932 paper ``Disintegration of Lithium by Swift Protons'' \cite{1932Coc01}.
Protons accelerated by a voltage between 125 and 400~kV bombarded a lithium target. Charged particle tracks were measured with a Shimizu expansion chamber. `` The brightness of the scintillations and the density of the tracks observed in the expansion chamber suggest that the particles are normal $\alpha$-particles. If this point of view turns out to be correct, it seems not unlikely that the lithium isotope of mass 7 occasionally captures a proton and the resulting nucleus of mass 8 breaks into two $\alpha$-particles, each of mass four and each with an energy of about eight million electron volts.'' This interpretation was confirmed by a measurement of two back-to-back $\alpha$-particles in coincidence reported only a few months later \cite{1932Coc02}.

\subsubsection*{$^9$Be}

In the 1921 article ``Anode Rays of Beryllium'' Thomson described the first observation of $^9$Be \cite{1921Tho01}. $^9$Be was detected in a mass spectrograph at Cambridge, England. ``A well-marked parabola was found corresponding to a single charge and an atomic weight 9.0$\pm$0.1 (Na=23). No second line was observed which could with certainty be attributed to beryllium, but the parabola at 9.0 was not so strong as that at 7.0 for lithium, and it is doubtful if one of a tenth the intensity could be observed.''

\subsubsection*{$^{10}$Be}

The existence of $^{10}$Be was shown by Oliphant et al. in 1935: ``Some Nuclear Transformations of Beryllium and Boron, and the Masses of the Light Elements'' \cite{1935Oli01}. The range of charged particles produced in the reaction of deuterons on beryllium and their deflection through magnetic and electric fields was measured. ``The proton group of range 26 cm can only be accounted for by assuming that a new isotope of beryllium of mass 10 is formed in the reaction $_4$Be$^9$ + $_1$H$^2 \to _4$Be$^{10}$ + $_1$h$^1$ + $\delta$, where the value of $\delta$ calculated from the observed range is 0.0050 mass units.''

\subsubsection*{$^{11}$Be}

$^{11}$Be was discovered by Nurmia and Fink in 1958 in ``New Isotope of Beryllium'' \cite{1958Nur01}. The University of Arkansas 400-kV Cockcroft-Walton accelerator was used to produce 14.8 MeV neutrons in the d-T reaction and $^{11}$Be was produced in the $^{11}$B(n,p) reaction. The decay curves were measured with $\beta$- and $\gamma$-scintillation counters and end-window GM tubes. ``The 14.1-second activity, which has not been reported previously, is assigned tentatively to a new isotope of beryllium, Be$^{11}$, from the B$^{11}$(n,p) reaction. From a series of measurements, a value of 14.1$\pm$0.3 seconds is derived for the half-life...'' This half-life agrees with the currently adopted value of 13.81(8)~s.

\subsubsection*{$^{12}$Be}

Poskanzer et al. discovered $^{12}$Be in 1966 in ``New Isotopes: $^{11}$Li, $^{14}$B, and $^{15}$B'' \cite{1966Pos01}. Uranium foils were bombarded with 5.3~GeV protons from the Berkeley Bevatron. Phosphorus-diffused silicon transmission detectors were used in a telescope consisting of an energy-loss, energy, and rejection detector to identify the isotopes. ``In fact, the assignment of an (11.4$\pm$0.5)-msec delayed neutron activity to $^{12}$Be on the assumption that $^{11}$Li was particle unstable must be re-examined... In any event, the particle stability of both $^{11}$Li and $^{12}$Be is established in the present data.'' The half-life quoted refers to a previous paper \cite{1965Pos02} and differs from the currently accepted value of 21.50(4)~ms by almost a factor of two.

\subsubsection*{$^{13}$Be}

An unbound resonance in $^{13}$Be was first reported in 1983 by Aleksandrov et al. in ``Observation of the isotope $^{13}$Be in the reaction $^{14}$C($^7$Li,$^8$B)'' \cite{1983Ale01}. An 82 MeV $^7$Li beam accelerated by the Kurchatov Institute isochronous cyclotron bombarded a self-supporting $^{14}$C target. $^{13}$Be was identified by measuring charged particles in a $\Delta$-E, transmission E and anti-coincidence energy telescope. ``...This permits one to conclude that the observed maximum in the reaction $^{14}$C($^7$Li,$^8$B) is due to the formation of the residual nucleus $^{13}$Be... On the basis of the calibration carried out, it was found that the mass defect of the $^{13}$Be nucleus is 35$\pm$0.5 MeV and consequently the isotope $^{13}$Be is unstable by 1.8 MeV against the decay $^{13}$Be $\to ^{12}$Be +n.''

\subsubsection*{$^{14}$Be}

In 1973 $^{14}$Be was discovered by Bowman et al. in ``Discovery of Two Isotopes, $^{14}$Be and $^{17}$B, at the Limits of Particle Stability'' \cite{1973Bow01}. A uranium target was bombarded with 4.8 GeV protons from the Berkeley Bevatron and fragments were identified by $\Delta$-E/E, and time-of-flight measurements in a silicon telescope. ``Two new isotopes, $^{14}$Be and $^{17}$B, were observed to be particle stable, and two others, $^{12}$Li and $^{16}$B, were shown to be particle unstable. The new isotope $^{17}$B recently had been predicted to be particle stable, but the observation of $^{14}$Be was surprising because it was thought to be unstable on the basis of both theoretical predictions and previous experimental results.'' An earlier report of the instability of $^{14}$Be \cite{1970Art03} was thus not confirmed.

\subsection{Boron}\vspace{0.0cm}

Thirteen boron isotopes are described including 2 stable, 1 proton-rich, 6 neutron-rich, and 2 proton- and 2 neutron-unbound resonances. At least the one-neutron unbound resonance $^{20}$B should be able to be observed in the future.

\subsubsection*{$^7$B}

McGrath et al. reported for the first time an unbound state of $^{7}$B in ``Unbound Nuclide $^7$B'' in 1967 \cite{1967McG01}. A carbon-backed $^{10}$B target was bombarded with a 50 MeV $^3$He beam from the Berkeley 88-in. cyclotron. Charged particles from the reaction $^{10}$B($^3$He,$^6$He)$^7$B were measured in two four-counter semiconductor telescopes. ``...the broad peak at about 29.8 MeV is attributed to the unbound ground state of $^7$B... The mass excess [$^{12}$C=0] of $^7$B was found to be 27.94$\pm$0.10 MeV.''

\subsubsection*{$^8$B}

$^8$B was discovered by Alvarez in the 1950 paper ``Three New Delayed Alpha-Emitters of Low Mass'' \cite{1950Alv01}. A 32 MeV proton beam from the Berkeley linear accelerator bombarded a proportional counter filled with $^{10}$BF$_3$ and delayed heavy particles were observed in the counter. ``Two new positron active isotopes, B$^8$ and Na$^{20}$, have been found to decay to excited states of Be$^8$ and Ne$^{20}$, which in turn decay ``instantaneously'' by alpha-emission. Their half-lives are 0.65$\pm$0.1 sec. and 1/4 sec., respectively. The half-life of $^8$B is consistent with the currently accepted value of 770(3)~ms.

\subsubsection*{$^9$B}

First evidence for the unbound nature of $^9$B was demonstrated by Haxby et al. in ``Threshold for the Proton-Neutron Reactions of Lithium, Beryllium, Boron, and Carbon'' in 1940 \cite{1940Hax01}. Protons up to 3.7 MeV from the Westinghouse pressure electrostatic generator bombarded beryllium targets and neutrons from the (p,n) charge exchange reaction were measured with a BF$_3$ ionization chamber. ``The observed energy difference (B$^9$$-$Be$^9$) would permit a positron radioactivity with maximum positron energy of 0.06 Mev or a K-electron capture... One may account for the absence of such activity in B$^9$ by the fact that B$^9$ is unstable with regard to dissociation into Be$^8$ + H$^1$... Hence we conclude that the Be$^9$(p,n)$^9$B reaction is immediately followed by the B$^9 \to$ He$^4$ + He$^4$ + H$^1$ disintegration.'' The extracted mass for $^9$B was 9.01600(13) amu.

\subsubsection*{$^{10,11}$B}

Aston discovered $^{10}$B and $^{11}$B in 1920 as reported in ``The constitution of the elements'' \cite{1920Ast02}. The isotopes were identified by measuring their mass spectra. ``Boron (atomic weight 10.9) is a complex element. Its isotopes are 10 and 11, satisfactorily confirmed by second-order lines at 5 and 5.5.''

\subsubsection*{$^{12}$B}

In 1935 Crane et al. observed $^{12}$B for the first time in ``The Emission of Negative Electrons from Boron Bombarded by Deuterons'' \cite{1935Cra01}. Deuterons accelerated by 500 kV bombarded a metal boron target and $^{12}$B was produced in the reaction $^{11}$B(d,p). Protons and electrons were detected in cloud chambers where the electron tracks were curved by a 1500 G magnetic field. ``The conclusion to be drawn from this is that B$^{12}$, in disintegrating, loses an amount of mass not less than that corresponding to the upper limit of energy of the electron spectrum. The large energy involved in the electron spectrum and the fact that only the mass difference between n$^1$ and H$^1$ is made use of, place this conclusion well beyond the limits of experimental error, provided only that the reactions assumed are correct.'' The measured half-life of 20~ms agrees with the currently accepted value of 20.20(2)~ms.

\subsubsection*{$^{13}$B}

In 1956, Allison et al. reported the first observation of $^{13}$B in ``Mass of B$^{13}$ from the Nuclear Reaction Li$^7$(Li$^7$,p)B$^{13}$'' \cite{1956All01}. A 1.61 MeV $^7$Li beam was accelerated by a Van de Graaff accelerator and bombarded a LiF target. Charged particles emitted at 90$^\circ$ were detected with a CsI(Tl) scintillating crystal. ``Using published mass values, it results that the Q-value of the new Li$^7$(Li$^7$,p)B$^{13}$ reaction is 5.97$\pm$0.03 Mev, giving B$^{13}$, presumably in its ground state, a value of (M$-$A) equal to 20.39$\pm$0.03 Mev, or a physical atomic weight of 13.02190$\pm$0.00003.'' A previous attempt to measure neutron activity from $^{13}$B as a delayed neutron emitter was unsuccessful \cite{1953Hub01}.

\subsubsection*{$^{14,15}$B}

Poskanzer et al. discovered $^{14}$B and $^{15}$B in 1966 in ``New Isotopes: $^{11}$Li, $^{14}$B, and $^{15}$B'' \cite{1966Pos01}. Uranium foils were bombarded with 5.3~GeV protons from the Berkeley Bevatron. Phosphorus-diffused silicon transmission detectors were used in a telescope consisting of an energy-loss, energy, and rejection detector to identify the isotopes. ``The predicted locations of the observed $^{14}$B and $^{15}$B peaks are indicated on the figure by arrows as are the expected positions of the neighboring isotopes $^9$C and $^{10}$C... Since $^{15}$B was predicted to be bound and $^{14}$B was expected to be marginally bound, the present observations of their existence were not unexpected.''

\subsubsection*{$^{16}$B}

Kalpakchieva et al. observed $^{16}$B for the first time as reported in the 2000 paper ``Spectroscopy of $^{13}$B, $^{14}$B, $^{15}$B and $^{16}$B using multi-nucleon transfer reactions'' \cite{2000Kal01}. A 336.6 MeV $^{14}$C accelerated by the heavy-ion accelerator at HMI-Berlin bombarded a $^{14}$C target and $^{16}$B was produced in the reaction $^{14}$C($^{14}$C,$^{12}$N). The ejectiles were detected with the Q3D magnetic spectrometer and the isotopes were identified by energy loss and time-of-flight measurements. ``Between the two $^{14}$B-states at 8.03 MeV and 10.15 MeV the lowest-lying $^{16}$B peak can be identified. It is found at a position which corresponds to a value of Q$_0$ = $-$48.38(6) MeV. Thus, a mass excess of 37.08(6) MeV can be determined for this $^{16}$B-peak.''

\subsubsection*{$^{17}$B}

In 1973 $^{17}$B was discovered by Bowman et al. in ``Discovery of Two Isotopes, $^{14}$Be and $^{17}$B, at the Limits of Particle Stability'' \cite{1973Bow01}. A uranium target was bombarded with 4.8 GeV protons from the Berkeley Bevatron and fragments were identified by $\Delta$-E/E, and time-of-flight measurements in a silicon telescope. ``Two new isotopes, $^{14}$Be and $^{17}$B, were observed to be particle stable, and two others, $^{12}$Li and $^{16}$B, were shown to be particle unstable. The new isotope $^{17}$B recently had been predicted to be particle stable, but the observation of $^{14}$Be was surprising because it was thought to be unstable on the basis of both theoretical predictions and previous experimental results.''

\subsubsection*{$^{18}$B}

An unbound state of $^{18}$B was first reported in 2010 by Spyrou et al. in ``First Evidence for a Virtual $^{18}$B Ground State'' \cite{2010Spy01}. A secondary beam of 62 MeV/u $^{19}$C produced by the Michigan State Coupled Cyclotron Facility and the A1900 fragment separator bombarded a beryllium target and $^{18}$B was produced in a one-proton knockout reaction. The excitation energy spectrum of $^{18}$B was reconstructed by measuring neutrons in coincidence with $^{17}$B fragments. ``An s-wave line shape was used to describe the experimental spectrum resulting in an upper limit for the scattering length of $-$50 fm which corresponds to a decay energy $<$10 keV. Observing an s-wave decay of $^{18}$B provides experimental verification that the ground state of $^{19}$C includes a large s-wave component.''

\subsubsection*{$^{19}$B}

Musser and Stevenson discovered $^{19}$B in 1984 in ``First Observation of the Neutron-Rich Isotope $^{19}$B'' \cite{1984Mus01}. The Berkeley Bevalac accelerated a beam of $^{56}$Fe to 670 A$\cdot$MeV which bombarded a beryllium target. Projectile fragments were measured and identified with the 0$^\circ$ spectrometer facility using a detector telescope consisting of a wire-chamber hodoscope, a front scintillator paddle, a set of threshold Cherenkov counters and a back scintillator paddle. ``Also shown in this figure are the predicted locations of $^{15}$B, $^{17}$B, and $^{19}$B. The previously observed particle instability of $^{16}$B and particle stability of $^{17}$B is apparent in this figure. In addition, our data indicate for the first time that $^{19}$B is particle stable, while $^{18}$B is unstable to prompt neutron emission.''

\subsection{Carbon}\vspace{0.0cm}

Fourteen carbon isotopes are described including 2 stable, 3 proton-rich, 8 neutron-rich, and 1 proton-unbound resonance. Carbon is the lightest element for which the properties of an unbound resonance of an isotope ($^{21}$C) which is lighter than another bound isotope ($^{22}$C) has not been studied yet. $^{21}$C had initially been reported as bound \cite{1981Ste01}, however this could not be confirmed \cite{1985Lan01}. $^{23}$C is probably the most difficult unbound nucleus delineating the neutron-dripline to be studied because of the location of the dripline at N=14 ranging from carbon to oxygen.

\subsubsection*{$^8$C}

$^8$C was discovered by Robertson et al. in the 1974 paper ``Highly Proton-Rich T$_z$ = $-$2 Nuclides: $^8$C and $^{20}$Mg'' \cite{1974Rob01}. Alpha-particles accelerated to 156 MeV by the J\"ulich isochronous cyclotron bombarded a natural carbon target and produced $^8$C in the reaction $^{12}$C($\alpha$,$^8$He). The $^8$He ejectiles were measured in a double-focusing magnetic analyzer and the energy-loss, energy, magnetic rigidity and time-of-flight were recorded. ``The measured mass excess of $^8$C is 35.30$\pm$0.20 MeV, and $^8$C is thus unbound.''

\subsubsection*{$^9$C}

$^9$C was first observed in 1964 by Cerny et al. in ``Completion of the mass-9 isobaric quartet via the three-neutron pickup reaction C$^{12}$(He$^3$,He$^6$)C$^9$'' \cite{1964Cer01}. A beam of 65 MeV $^3$He from the Berkeley 88-in. variable-energy cyclotron impinged on a $^{12}$C target. $^9$C was produced in the reaction C$^{12}$(He$^3$,He$^6$)C$^9$ and identified with a $\Delta$E-E semiconductor counter telescope. ``The mass excess of C$^9$ on the C$^{12}$ scale was determined to be 28.95$\pm$0.15 MeV; hence, as expected, C$^9$ is stable with respect to proton emission.''

\subsubsection*{$^{10}$C}

The first observation of $^{10}$C was reported by Sherr et al. in 1949: ``Radioactivity of C$^{10}$ and O$^{14}$'' \cite{1949She01}. Boron metal or boron compounds were bombarded by 17 MeV protons from the Princeton cyclotron. Radioactive gases from the target were collected and chemically separated. $^{10}$C was formed in the (p,n) charge exchange reaction and positrons and $\gamma$-rays were detected. ``In addition to the well-known 20.5-min. activity of C$^{11}$ produced by a (p,n) reaction in B$^{11}$, a new period of 19.1 sec. was found.'' This half-life of 19.1(8)~s agrees with the presently accepted value of 19.290(12)~s. A previous reported half-life for $^{10}$C of 8.8(8)~s \cite{1940Del01} could not be confirmed and Sherr et al. suggested that it must have been due to impurities in the boron powder.

\subsubsection*{$^{11}$C}

In 1934 Crane and Lauritsen reported the discovery of $^{11}$C in ``Radioactivity from Carbon and Boron Oxide Bombarded with Deutons and the Conversion of Positrons into Radiation'' \cite{1934Cra01}. A B$_2$O$_3$ target was bombarded with deuterons accelerated by 900 kV and $^{11}$C was identified by measuring the decay curve and $\gamma$-rays. ``The B$_2$O$_3$ target gave an effect which was somewhat smaller than that obtained from carbon, and decreased at a rate corresponding to a half life of about 20 minutes... Although this target has not yet been tested in a cloud chamber, the activity here concerned seems to be of the same nature as that of carbon, and by analogy we suppose that the transmutations are $_5$B$^{10}$ + $_1$H$^2 \to _6$C$^{11}$ + $_0$n$^1$, $_6$C$^{11} \to _5$B$^{11}$ + (+e).''

\subsubsection*{$^{12}$C}

The 1919 paper ``The Constitution of the Elements'' by Aston can be considered the discovery of $^{12}$C \cite{1919Ast01}. $^{12}$C was identified using the positive-ray mass spectrograph in Cambridge, England. ``Of the elements involved hydrogen has yet to be investigated; carbon and oxygen appear, to use the terms suggested by Paneth, perfectly ``pure''... A fact of the greatest theoretical interest appears to underlie these results, namely, that of more than forty different values of atomic and molecular mass so far measured, all, without a single exception, fall on whole numbers, carbon and oxygen being taken as 12 and 16 exactly, and due allowance being made for multiple charges.''

\subsubsection*{$^{13}$C}

$^{13}$C was first observed in 1929 by King and Birge in ``An Isotope of Carbon, Mass 13'' \cite{1929Kin01}. The isotope was identified in the Swan spectrum of carbon in a vacuum electric furnace. ``Although with the present available data the fact may have no significance, it is interesting to note that this small discrepancy may be canceled by assuming 12.0000 and 13.0026 for the two masses... The present evidence seems however fully sufficient to establish the existence of an isotope of carbon, of mass 13.'' It should be noted that the same article was simultaneously submitted to Nature, where it appeared 12 days later \cite{1929Kin02}.

\subsubsection*{$^{14}$C}

The first identification of $^{14}$C was presented in ``The disintegration of Nitrogen by Neutrons'' by Bonner and Brubaker in 1936 \cite{1936Bon01}. The neutron capture reaction on $^{14}$N was studied and reaction products were measured in a cloud chamber. In addition to the reaction $^{14}$N(n,$\alpha$)$^{11}$B (1), the two possible reactions with single-charged ejectiles $^{14}$N(n,d)$^{13}$C (2) and $^{14}$N(n,p)$^{14}$C (3) were considered. ``According to calculations from Bethe's masses, reaction (2) is endothermic by 4.7 MeV. Thus we must turn to reaction (3) to explain the singly charged particles. $_7$N$^{14}$ + $_0$n$^1 \to _6$C$^{14}$ + $_1$H$^1$. The C$^{14}$ would probably be radioactive, going into N$^{14}$ with the emission of an electron. However, such a radioactive C$^{14}$ has not been observed.'' Previously, Kurie had mentioned the possibility that neutron capture on $^{14}$N could lead to the formation of $^{14}$C, however he could not distinguish the other possibilities of $^{14}$N(n,d)$^{13}$C or $^{14}$N(n,t)$^{12}$C \cite{1934Kur01}. Also, Miller et al. had speculated that $^{14}$C had been produced in the reaction $^{11}$B($\alpha$,p)$^{14}$C, however, they stated ``...until there is more evidence on this point it seems reasonable to assume that all the four energy changes are connected with the B$^{10}$ isotope'' \cite{1934Mil01}.

\subsubsection*{$^{15}$C}

Hudspeth et al. described the discovery of $^{15}$C in the 1950 paper ``Production of C$^{15}$'' \cite{1950Hud01}. Deuterons accelerated to 2.4 MeV bombarded a BaCO$_3$ target (enriched to 40\% $^{14}$C) and $^{15}$C was produced in the (d,p) reaction. Decay curves of the resulting $\beta$-ray activity were measured. ``All of the sets of data indicated a half-life of 2.4 seconds, with an estimated error of about 0.3 second.'' This half-life agrees with the currently adopted value of 2.449(5)~s.

\subsubsection*{$^{16}$C}

$^{16}$C was discovered in 1961 by Hinds et al. in ``New Isotope of Carbon: C$^{16}$'' \cite{1961Hin01}. The Aldermaston electrostatic generator was used to accelerate tritons to 6 MeV which bombarded a $^{14}$C target. $^{16}$C was produced in the (t,p) reaction and the protons were analyzed with a broad-range magnetic spectrograph. In addition, the half-life was determined by measuring $\beta$-delayed neutrons. ``The energy of the incident triton beam was determined in terms of the Q value of the C$^{12}$(t,p)C$^{14}$ reaction and from this the Q value of the C$^{14}$(t,p)C$^{16}$ reaction was calculated to be $-3.014\pm$0.016 Mev. This is consistent with a C$^{16}$ mass excess of 13.694$\pm$0.017 Mev and a mass of 16.014702$\pm$0.000017 mass units, both referred to the scale on which the mass excess of C$^{12}$ is zero.'' The measured half-life of 0.74(3)~s agrees with the presently accepted value of 747(8)~ms.

\subsubsection*{$^{17}$C}

In 1968, Poskanzer reported the first observation of $^{17}$C in ``Observation of the new isotope $^{17}$C using a combined time-of-flight particle-identification technique'' \cite{1968Pos01}. The Berkeley bevatron accelerated protons to 5.5 GeV which bombarded a uranium metal target. $^{17}$C was identified in a five-detector telescope measuring energy-loss, energy, and time-of-flight. ``The particle spectrum obtained during 5 days of data collection showed clear evidence for a $^{17}$C peak... Almost all the events which fell in the $^{17}$C region of the particle spectrum also fell in the mass 17 region of the mass spectrum, thus clearly providing the existence of $^{17}$C.''

\subsubsection*{$^{18}$C}

Artukh et al. reported the first identification of $^{18}$C in 1969 in ``New isotopes $^{22}$O, $^{20}$N and $^{18}$C produced in transfer reactions with heavy ions'' \cite{1969Art01}. $^{18}$O was accelerated by the Dubna 310 cm heavy ion cyclotron to 122 MeV and bombarded a metallic $^{232}$Th target. $^{18}$C was produced in transfer reactions and identified by energy-loss, energy and magnetic rigidity measurements in the focal plane of a magnetic analyzer. ``Along with the known isotopes some new ones have been found: $^{22}$O (about 100 events), $^{20}$N (about 60 events) and $^{18}$C (about 50 events).''

\subsubsection*{$^{19}$C}

In 1974 $^{19}$C was observed by Bowman et al. in ``Detection of neutron-excess isotopes of low-Z elements produced in high-energy nuclear reactions'' \cite{1974Bow01}. A uranium target was bombarded with 4.8 GeV protons from the Berkeley Bevatron and fragments were identified by $\Delta$-E vs E, and time-of-flight measurements in a silicon telescope. ``The mass-yield histograms show the existence of all previously known particle-stable nuclei in this region except for $^9$C. In addition, two new nuclei, $^{14}$Be and $^{17}$B, are clearly observed and the stability of $^{19}$C can be confirmed.'' Bowman et al. do not take credit for the discovery of $^{19}$C referring to a previous conference proceeding \cite{1970Rai01}. This reference does not represent the discovery of $^{14}$Be and $^{17}$B, because, as the manuscript states, the first observation of these isotopes had been reported in a previous brief report \cite{1973Bow01}.

\subsubsection*{$^{20}$C}

$^{20}$C was discovered by Stevenson and Price in the 1981 paper ``Production of the neutron-rich nuclides $^{20}$C and $^{27}$F by fragmentation of 213 MeV/nucleon $^{48}$Ca'' \cite{1981Ste01}. $^{48}$Ca at 213 MeV/nucleon from the Berkeley Bevatron was fragmented on a beryllium target. The fragments were focussed on a stack of Lexan plastic track detectors in the zero-degree magnetic spectrometer. ``There is clear evidence for the first observation of $^{20}$C ($\sim$40 counts) and $^{27}$F ($\sim$20 counts).''

\subsubsection*{$^{22}$C}

The discovery of $^{22}$C was reported in 1986 by Pougheon et al. in ``First Observation of the Exotic Nucleus $^{22}$C'' \cite{1986Pou01}. A 44 MeV/u $^{40}$Ar beam was fragmented on a tantalum target at GANIL. The fragments were measured with the triple-focusing magnetic spectrometer LISE and identified by measuring energy-loss, energy and time-of-flight. ``The nuclei with a ratio of A/Z = 3 (i.e. $^{15}$B, $^{18}$C, $^{21}$N) are clearly visible forming a line of constant time of flight. This, together with the observed absence in the plot of isotopes known to be particle unstable ($^{16}$B) ensures an unambiguous identification of the observed isotopes. In this plot the $^{19}$B and the $^{22}$C isotopes are observed without any surrounding background event.''

\subsection{Nitrogen}\vspace{0.0cm}

Fourteen nitrogen isotopes are described including 2 stable, 2 proton-rich, 8 neutron-rich, and 2 proton-unbound resonances. The one-neutron unbound resonance of $^{24}$N should still be able to be observed in the future.

\subsubsection*{$^{10}$N}

An unbound resonance in $^{10}$N was observed in 2002 by L\'epine-Szily et al. in ``Observation of the particle-unstable nucleus $^{10}$N'' \cite{2002Lep01}. A 30 MeV/nucleon $^{14}$N beam was accelerated at GANIL and bombarded an isotopically enriched $^{10}$B target. $^{10}$N was produced in the double-charge-exchange reaction $^{10}$B($^{14}$N,$^{14}$B) identified by measuring the ejectiles in the high-precision spectrometer SPEG. ``The peak observed in the subtracted spectrum has statistical significance of 4.0~$\sigma$... The peak is a broad structure and was fitted by an l=0 resonance at E$_r$ = 2.6(4) MeV. This resonance energy corresponds to a mass excess of 38.8(4) MeV for the $^{10}$N$_{g.s.}$, very close to the Audi-Wapstra estimation, which is about 38.5(4) MeV.''

\subsubsection*{$^{11}$N}

Benenson et al. discovered $^{11}$N in 1974 in ``T =$\frac{3}{2}$ states in mass-11 nuclei'' \cite{1974Ben01}. A $^{14}$N gas target was bombarded with a 70 MeV $^3$He beam at Michigan State University and $^{11}$N was produced in the ($^3$He,$^6$He) transfer reaction. $^{11}$N was identified in a magnetic spectrograph by measuring the time-of-flight in combination with a magnetic analysis. ``The kinematic effects in the $^{14}$N($^3$He,$^6$He)$^{11}$N reaction are quite marked even at forward angles, and therefore the observation of the same width and energy of a peak at all three angles is a strong indication that a $^{11}$N state is being studied... The level parameters for the peak near channel 25 are mass excess = 25.23$\pm$0.10 MeV and $\Gamma$ = 740$\pm$100 keV.''

\subsubsection*{$^{12}$N}

$^{12}$N was discovered in 1949 by Alvarez as reported in ``Nitrogen 12'' \cite{1949Alv02}. Protons accelerated to 30 MeV by a linear accelerator at Berkeley bombarded a carbon target. $^{12}$N was produced in the (p,n) charge exchange reaction and identified by recording the positron activity in coincidence with a pair of trays of Geiger counters. ``N$^{12}$ is shown to have a half-life of 12.5$\pm$1 milliseconds, and a positron upper limit of 16.6$\pm$0.2 Mev... The mass of N$^{12}$ is 12.0228$\pm$0.00015, and the beta-transition is allowed.'' This half-life is close to the currently accepted value of 11.000(16)~ms.

\subsubsection*{$^{13}$N}

Curie and Joliot presented first evidence of $^{13}$N in 1934 in ``Un nouveau type de radioactivit\'e'' \cite{1934Cur01}. Aluminum, boron, and magnesium samples were irradiated by polonium $\alpha$-particles and their activities were measured with a Geiger M\"uller counter as a function of time. The half-life of the activity created in aluminum was 3.25~min and the reaction $^{27}_{13}$Al + $^4_2$He = $^{30}_{15}$P + $^1_0$n was proposed: ``On obtient un r\'esultat analogue avec le bore et le magn\'esium mais les p\'eriodes de d\'ecroissance sont diff\'erentes, 14 minutes pour le bore et 2 minutes 30 secondes pour le magn\'esium... Une r\'eaction analogue pourrait \^etre envisag\'ee pour le bore et le magn\'esium, les noyaux instables \'etant $^{13}_7$N et $^{27}_{14}$Si.`` [We obtain a similar result with boron and magnesium, but the decay periods are different, 14 minutes for boron and 2 minutes and 30 seconds for magnesium ... A similar reaction could be considered for boron and magnesium with the unstable nuclei $^{13}_7$N and $^{27}_{14}$Si.] The half-life of about 14~min. for $^{13}$N is close to the currently adopted value of 9.965(4)~min.

\subsubsection*{$^{14}$N}

In 1920 $^{14}$N was first measured by Aston in ``The Constitution of the Elements'' \cite{1920Ast01}. The isotope was identified in a mass spectrometer at Cambridge, England. ``Nitrogen is apparently a `pure' element, its doubly charged atom being 7 exactly.''

\subsubsection*{$^{15}$N}

Naude reported the first observation of $^{15}$N in the 1929 paper ``An isotope of nitrogen, mass 15'' \cite{1929Nau01}. The continuous light of a hydrogen lamp was used to illuminate NO gas and the absorption spectra were measured with an E$_1$ Hilger spectrograph. ``The mean values of measurements on five different plates give the following wave-lengths for the isotope heads: Q$_1$ heads: 2155.227(N$^{14}$O$^{17}$), 2155.730(N$^{15}$O$^{16}$), 2156.753(N$^{14}$O$^{18}$). P$_1$ heads: 2156.493(N$^{14}$O$^{17}$), 2156.982(N$^{15}$O$^{16}$), 2157.976(N$^{14}$O$^{18}$)... At this stage it can only be said that N$^{15}$O$^{16}$ is about as abundant as N$^{14}$O$^{18}$.''

\subsubsection*{$^{16}$N}

First evidence of the existence of $^{16}$N was presented by Harkins et al. in the 1933 paper ``Disintegration of Fluorine Nuclei by Neutrons and the Probable Formation of a New Isotope of Nitrogen (N$^{16}$)'' \cite{1933Har01}. Neutrons irradiated a mixture of 30\% difluor-dichlor-methane and 70\% helium in a Wilson cloud chamber and 3200 pairs of photographs were taken. Ten photographs exhibited nuclear disintegrations. ``The momentum and energy relations, together with other evidence, indicate that most, and probably all of these are disintegrations of fluorine nuclei... Thus for disintegrations in which the neutron is captured the reaction may be considered to be represented by F$_1^{19}$ + n$_1^1 \to$ F$_2^{20} \to$ N$_2^{16}$ + He$_0^4$, in which the subscripts represent the isotopic numbers and the superscripts the atomic masses.''

\subsubsection*{$^{17}$N}

In 1949 Alvarez discovered $^{17}$N in ``N$^{17}$, A Delayed Neutron Emitter'' \cite{1949Alv01}. A water solution of NH$_4$F was bombarded with 190 MeV deuterons from the Berkeley 184 inch cyclotron. $^{17}$N was formed in the reaction $^{19}$F(d,$\alpha$) and delayed neutron activity was counted with a BF$_3$ chamber. ``The conclusion, therefore, comes largely from a process of elimination, but it seems quite conclusive. The nuclear evidence supports it strongly and there is little question that even without any chemical work at all, the activity would have been assigned to N$^{17}$.'' Further chemical and physical evidence supported this first conclusion. The observed half-life of 4.2~s agrees with the present value of 4.173(4)~s.

\subsubsection*{$^{18}$N}

$^{18}$N was observed by Chase et al. in 1964 in ``New isotope of nitrogen: N$^{18}$'' \cite{1964Cha01}. Neutrons of 19 MeV from the reaction T(d,n)$^4$He produced by the Lockheed 3.5 MeV Van de Graaff accelerator irradiated a water sample containing enriched $^{18}$O. $^{18}$N was formed in the (n,p) charge exchange reaction. Gamma- and $\beta$-rays of the activated samples were measured with NaI(Tl) scintillators and a plastic scintillator $\Delta$E-E telescope, respectively. ``This decay curve, along with many other similar curves, gives a weighted average half-life for N$^{18}$ of 0.63$\pm$0.03 sec.'' This half-life agrees with the presently accepted value of 622(9)~ms.

\subsubsection*{$^{19}$N}

Thomas et al. first reported the observation of $^{19}$N in ``New isotopes $^{19}$N and $^{21}$O, produced in high-energy nuclear reactions'' in 1968 \cite{1968Tho01}. The Princeton-Pennsylvania Accelerator was used to bombard a gold target with 3 GeV protons and the resulting fragments were identified by energy-loss, energy and time-of-flight measurements with a telescope consisting of four surface barrier detectors. ``There is clear evidence in their spectrum for the isotopes $^{11}$Li and $^{14}$Be reported by Poskanzer et al. and for the new isotopes $^{19}$N and $^{21}$O.''

\subsubsection*{$^{20}$N}

Artukh et al. reported the first identification of $^{20}$N in 1969 in ``New isotopes $^{22}$O, $^{20}$N and $^{18}$C produced in transfer reactions with heavy ions'' \cite{1969Art01}. $^{18}$O was accelerated by the Dubna 310 cm heavy ion cyclotron to 122 MeV and bombarded a metallic $^{232}$Th target. $^{20}$N was produced in transfer reactions and identified by energy-loss, energy and magnetic rigidity measurements in the focal plane of a magnetic analyzer. ``Along with the known isotopes some new ones have been found: $^{22}$O (about 100 events), $^{20}$N (about 60 events) and $^{18}$C (about 50 events).''

\subsubsection*{$^{21}$N}

In 1970 $^{21}$N was discovered by Artukh et al. in ``New isotopes $^{21}$N, $^{23}$O, $^{24}$O and $^{25}$F, produced in nuclear reactions with heavy ions'' \cite{1970Art02}. A metallic $^{232}$Th target was bombarded with a 174 MeV $^{22}$Ne beam from the 310 cm heavy ion cyclotron at Dubna, Russia. The reaction products were identified in a $\Delta$E-E semiconductor telescope at the focal plane of a magnetic spectrometer. ``[The figure] shows that apart from a number of already known isotopes, four new isotopes: $^{21}$N (about 60 events), $^{23}$O (about 130 events), $^{24}$O (about 30 events) and $^{25}$F (about 40 events) have been obtained.''

\subsubsection*{$^{22}$N}

The first observation of $^{22}$N was reported by Westfall et al. in ``Production of neutron-rich nuclides by fragmentation of 212-MeV/amu $^{48}$Ca'' in 1979 \cite{1979Wes01}. $^{48}$Ca ions (212 MeV/nucleon) from the Berkeley Bevalac were fragmented on a beryllium target. The fragments were selected by a zero degree spectrometer and identified in a telescope consisting of 12 Si(Li) detectors, 2 position-sensitive Si(Li) detectors, and a veto scintillator. ``There is clear evidence for the particle stability of $^{22}$N, $^{26}$F, $^{33,34}$Al, $^{37,38,39}$Si, $^{40,41,42}$P, $^{41,42,43,44}$Si, and $^{44,45}$Cl with more than ten counts in each case.''

\subsubsection*{$^{23}$N}

The discovery of $^{23}$N was reported in 1985 by Langevin et al. in ``Production of neutron-rich nuclei at the limits of particle stability by fragmentation of 44 MeV/u $^{40}$Ar projectiles'' \cite{1985Lan01}. A 44 MeV/u $^{40}$Ar beam was fragmented on a tantalum target at GANIL. The fragments were measured with the triple-focusing magnetic spectrometer LISE and identified by measuring energy-loss, energy and time-of-flight. ``The first observation of $^{23}$N, $^{29}$Ne and $^{30}$Ne and their particle bound character results clearly from the mass histograms of [the figures].''

\subsection{Oxygen}\vspace{0.0cm}

Fourteen oxygen isotopes are described including 3 stable, 3 proton-rich, 6 neutron-rich, and 1 proton- and 1 neutron-unbound resonance. The two-neutron unbound resonance of $^{26}$O should be observed in the future due to the special interest in the vanishing of the N=20 and the appearance of the N=16 shell in neutron-rich nuclei. Oxygen is currently the heaviest element for which all particle-bound isotopes have been discovered. Thus the location of the neutron dripline is unknown beyond oxygen. $^{25}$O is also the heaviest neutron-unbound nucleus which has been characterized at the present time.

\subsubsection*{$^{12}$O}

$^{12}$O was first observed by KeKelis et al. in the 1978 paper ``Masses of the unbound nuclei $^{16}$Ne, $^{15}$F, and $^{12}$O'' \cite{1978KeK01}. An oxygen gas target was bombarded with 117 MeV $\alpha$ particles from the Berkeley 88-inch cyclotron. $^{12}$O was produced in the transfer reaction $^{16}$O($^4$He,$^8$He) and identified by measuring the ejectiles in a quadrupole-sextupole-dipole (QSD) spectrometer. ``...there is evidence for a group of seven counts near the location of the IMME prediction for the $^{12}$O ground state as well as five counts which could represent transitions to a first excited state... This Q value implies a mass excess of 32.10$\pm$0.12 MeV for $^{12}$O.''

\subsubsection*{$^{13}$O}

In ``Observation of delayed proton radioactivity'' Barton et al. implied the observation of $^{13}$O for the first time in 1963 \cite{1963Bar01}. An air target was bombarded with 97 MeV protons from the McGill Synchrocyclotron. $^{13}$O was identified by the observation of $\beta$-delayed protons in a silicon junction particle detector. ``The hypothesis that the decay of the nuclide (2k+2,2k-1) will dominate the delayed proton spectrum from targets of both element (2k+2) and element (2k+1) seems to be verified. In particular, by following very reasonable rules for predicting proton lines, all the observed lines are accounted for and all those predictions based on known level properties are borne out. The existence of Mg$^{21}$, Ne$^{17}$, and O$^{13}$ is assumed since it seems fairly certain that proton lines in the decay of each have been observed.'' The half-life was subsequently measured by McPherson et al. \cite{1965McP01} who acknowledged the tentative observation by Barton et al.

\subsubsection*{$^{14}$O}

The first observation of $^{14}$O was reported by Sherr et al. in 1949: ``Radioactivity of C$^{10}$ and O$^{14}$'' \cite{1949She01}. Nitrogen gas and nitrogen compounds were bombarded by 17 MeV protons from the Princeton cyclotron. Radioactive gases from the target were collected and chemically separated. $^{14}$O was formed in the (p,n) charge-exchange reaction and positrons and $\gamma$-rays were detected. ``A new activity, O$^{14}$, has been produced from N$^{14}$ by a (p,n) reaction and is found to decay with a half-life of 76.5$\pm$2 sec. by the emission of 1.8$\pm$0.1 Mev positrons and a 2.3 Mev gamma-ray.'' This half-life is consistent with the presently accepted value of 70.598(18)~s.

\subsubsection*{$^{15}$O}

In 1934 Livingston and McMillan discovered $^{15}$O in ``The production of radioactive oxygen'' \cite{1934Liv01}. At Berkeley, deuterons accelerated by 2 MV bombarded nitrogen gas to produce $^{15}$O. The $\gamma$-ray emission from the positron annihilation was measured and the activated sample was chemically separated. ``The probable nuclear reactions are: $_7$N$^{14}$ + $_1$H$^2 \to _8$O$^{15}$ + $_0$n$^1$, $_8$O$^{15} \to _7$N$^{15}$ + $_1$e$^+$. The neutrons given by [the] reaction were found to be produced in the activating process in about the expected numbers.'' The observed half-life of 126~s is consistent with the present value of 122.24(16)~s.

\subsubsection*{$^{16}$O}

The 1919 paper ``The Constitution of the Elements'' by Aston can be considered the discovery of $^{16}$O \cite{1919Ast01}. $^{16}$O was identified using the positive-ray mass spectrograph in Cambridge, England. ``Of the elements involved hydrogen has yet to be investigated; carbon and oxygen appear, to use the terms suggested by Paneth, perfectly ``pure''... A fact of the greatest theoretical interest appears to underlie these results, namely, that of more than forty different values of atomic and molecular mass so far measured, all, without a single exception, fall on whole numbers, carbon and oxygen being taken as 12 and 16 exactly, and due allowance being made for multiple charges.''

\subsubsection*{$^{17}$O}

Blackett identified $^{17}$O for the first time in 1925 in the paper ``The Ejection of Protons from Nitrogen Nuclei, Photographed by the Wilson Method'' \cite{1925Bla01}. The tracks of alpha-particles from a thorium B + C source ($^{212}$Pb and $^{212}$Bi) in a condensation chamber were photographed. ``But amongst these normal forks due to elastic collisions, eight have been found of a strikingly different type... These eight tracks undoubtedly represent the ejection of a proton from a nitrogen nucleus. It was to be expected that a photograph of such an event would show an alpha-track branching into three... These eight forks however branch only into two... As no such track exists the alpha-particle cannot escape. In ejecting a proton from a nitrogen nucleus the alpha-particle is therefore itself bound to the nitrogen nucleus. The resulting new nucleus must have a mass 17, and, provided no electrons are gained or lost in the process, an atomic number of 8... It ought therefore to be an isotope of oxygen.''

\subsubsection*{$^{18}$O}

The first identification of $^{18}$O was reported by Giauque and Johnston in 1929 in  ``An isotope of oxygen, mass 18. Interpretation of the atmospheric absorption bands'' \cite{1929Gia01}. Giauque and Johnston reinvestigated the absorption band data of atmospheric oxygen by Dieke and Babcock \cite{1927Die01}. ``It occurred to us that the A' band might result from an isotope of oxygen and we have found that it is fully explained as originating from an oxygen molecule consisting of an atom of mass 16 combined with an atom of mass 18. Such an isotope has not previously been observed but its existence in small amount has certainly not been disproved. Since it is practically the same mass as water, it might easily be misinterpreted in a mass spectrograph.'' The paper was submitted on January 14, 1929 and published on May 6, 1929. On March 2, 1929 Giauque and Johnston published a shorter version of the results in Nature \cite{1929Gia02}. No submission date is given and it is assumed that it was submitted after the longer paper.

\subsubsection*{$^{19}$O}

In 1936 Nahmias and Walen identified $^{19}$O for the first time in ``Sur quelques radio\'el\'ements artificiels'' \cite{1936Nah01}. Neutrons from a Rn-Be source irradiated a LiF sample. Activities of 8.4(1)~s and 31(1)~s were observed which could originate from the following three reactions: (1) $^{19}$F + n $\to ^{20}$F, (2) $^{19}$F + n $\to ^{19}$O + $^1$H, and (3) $^{19}$F + n $\to ^{16}$N + $^4$He. ``La p\'eriode de 31 secondes serait alors due \`a la r\'eaction (2) qui est certainement moins fr\'equente que (3).'' [The period of 31 seconds would be due to reaction (2) which is certainly less common than (3)]. The half-life is close to the currently adopted value of 26.464(9)~s.

\subsubsection*{$^{20}$O}

The discovery of $^{20}$O was described in the 1959 paper ``Oxygen-20'' by Jarmie and Silbert \cite{1959Jar01}. An enriched $^{18}$O gas target was bombarded with 2.6 MeV tritons from an electrostatic accelerator and $^{20}$O was formed in the (t,p) reaction. The ejectiles were analyzed in a double-focussing magnetic spectrometer and detected in a CsI crystal scintillation spectrometer. ``Two proton groups associated with O$^{20}$ were observed. These are assumed to be due to the ground state and first excited state of O$^{20}$. The Q for the reaction O$^{18}$(t,p)O$^{20}$ is 3.12$\pm$0.04 Mev. This then gives a preliminary value for the mass of O$^{20}$ of 20.01036$\pm$0.00004 amu or a mass excess of 9.65$\pm$0.04 Mev.''

\subsubsection*{$^{21}$O}

Thomas et al. first reported the observation of $^{21}$O in ``New isotopes $^{19}$N and $^{21}$O, produced in high-energy nuclear reactions'' in 1968 \cite{1968Tho01}. The Princeton-Pennsylvania Accelerator was used to bombard 3 GeV protons on a gold target and the resulting fragments were identified by energy-loss, energy and time-of-flight measurements with a telescope consisting of four surface barrier detectors. ``There is clear evidence in their spectrum for the isotopes $^{11}$Li and $^{14}$Be reported by Poskanzer et al. and for the new isotopes $^{19}$N and $^{21}$O.''

\subsubsection*{$^{22}$O}

Artukh et al. reported the first identification of $^{22}$O in 1969 in ``New isotopes $^{22}$O, $^{20}$N and $^{18}$C produced in transfer reactions with heavy ions'' \cite{1969Art01}. $^{18}$O was accelerated by the Dubna 310 cm heavy ion cyclotron to 122 MeV and bombarded a metallic $^{232}$Th target. $^{22}$O was produced in transfer reactions and identified by energy-loss, energy and magnetic rigidity measurements in the focal plane of a magnetic analyzer. ``Along with the known isotopes some new ones have been found: $^{22}$O (about 100 events), $^{20}$N (about 60 events) and $^{18}$C (about 50 events).''

\subsubsection*{$^{23,24}$O}

In 1970 $^{23}$O and $^{24}$O were discovered by Artukh et al. in ``New isotopes $^{21}$N, $^{23}$O, $^{24}$O and $^{25}$F, produced in nuclear reactions with heavy ions'' \cite{1970Art02}. A metallic $^{232}$Th target was bombarded with a 174 MeV $^{22}$Ne beam from the 310 cm heavy ion cyclotron at Dubna, Russia. The reaction products were identified in a $\Delta$E-E semiconductor telescope at the focal plane of a magnetic spectrometer. ``[The figure] shows that apart from a number of already known isotopes, four new isotopes: $^{21}$N (about 60 events), $^{23}$O (about 130 events), $^{24}$O (about 30 events) and $^{25}$F (about 40 events) have been obtained.''

\subsubsection*{$^{25}$O}

An unbound state of $^{25}$O was first reported by Hoffman et al. in 2008 in ``Determination of the N = 16 Shell Closure at the Oxygen Drip Line'' \cite{2008Hof01}. A secondary beam of 85 MeV/u $^{26}$F produced by the Michigan State Coupled Cyclotron Facility and the A1900 fragment separator bombarded a beryllium target and $^{25}$O was produced in a one-proton knockout reaction. The excitation energy spectrum of $^{25}$O was reconstructed by measuring neutrons in coincidence with $^{24}$O fragments. ``A resonance energy of E$_{decay}$ = 770$^{+20}_{-10}$ keV and a width of $\Gamma$ = 172(30) keV were the best fit to the data... The determination of the mass of $^{25}$O from the resonance energy is dependent on the mass of $^{24}$O. A recent experiment on neutron-rich nuclei remeasured the mass of $^{24}$O, adopting a mass excess of 18600(100) keV, 470 keV below the currently accepted value of 19070(240) keV of the atomic mass evaluation (AME). The $^{25}$O mass excess is then 27440(110) keV and 27910(245) keV for [the new $^{24}$O mass data] and the AME, respectively.'' An initial report that $^{25}$O might be bound \cite{1981Ste01} could not be confirmed \cite{1985Lan01}.

\subsection{Fluorine}\vspace{0.0cm}

The observation of 16 fluorine isotopes has been reported so far, including 1 stable, 2 proton-rich, 10 neutron-rich, and 3 proton-unbound resonances. The one-neutron unbound resonances of $^{28}$F, $^{30}$F, and $^{32}$F should be able to be observed in the future. In addition, $^{33}$F still might be particle-stable.

\subsubsection*{$^{14}$F}

In 2010, Goldberg et al. reported the discovery of $^{14}$F in ``First observation of $^{14}$F'' \cite{2010Gol01}. A secondary 31 MeV/u $^{13}$O beam produced by in-flight separation at the Texas A\&M University Cyclotron was used to bombard a methane gas target. Proton-unbound states of $^{14}$F were populated by elastic scattering on hydrogen. ``The ground state and several low-lying excited states in $^{14}$F were observed and spin/parity assignments were made.''

\subsubsection*{$^{15}$F}

$^{15}$F was first observed by KeKelis et al. in the 1978 paper ``Masses of the unbound nuclei $^{16}$Ne, $^{15}$F, and $^{12}$O'' \cite{1978KeK01}. A neon gas target was bombarded with 87.8 MeV $^3$He particles from the Berkeley 88-inch cyclotron. $^{15}$F was produced in the transfer reaction $^{20}$Ne($^3$He,$^8$Li) and identified by measuring the ejectiles in a quadrupole-sextupole-dipole (QSD) spectrometer. ``The observed Q value of the ground state transition of $-$29.73$\pm$0.18 MeV corresponds to a mass excess of 16.67$\pm$0.18 MeV for $^{15}$F.'' It should be mentioned that $^{15}$F was observed independently by Benenson et al. \cite{1978Ben01}, submitted only a month later and published in the same issue, following the paper by KeKelis et al.. KeKelis acknowledges that ``A preliminary experiment by the MSU group at 75 MeV incident energy had produced evidence for the location of the $^{15}$F ground state.''

\subsubsection*{$^{16}$F}

Bryant et al. discovered $^{16}$F in ``(He$^{3}$,n) Reactions on Various Light Nuclei'' in 1964 \cite{1964Bry01}. 25 MeV $^3$He ions from the Los Alamos variable energy cyclotron bombarded a nitrogen gas target and the excitation energy spectrum of $^{16}$F was extracted by measuring neutron energies with a bubble chamber. ``The Reaction $^{14}$N($^3$He,$n$)$^{16}$F: Resolution of the ground state was not possible from our data, in light of the low lying excited states expected to be present from analogy with those found in N$^{16}$, the bimirror of F$^{16}$. The levels at 0.88 and 1.26 reported by Bonner et al. are not clearly in evidence.'' The reference to Bonner corresponds to conference proceeding \cite{1960Bon01}.

\subsubsection*{$^{17}$F}

``An artificial radioelement from nitrogen'' reported the first observation of $^{17}$F by Wertenstein in 1934 \cite{1934Wer01}. Alpha particles from a radon source bombarded nitrogen gas and the subsequent activity was measured with a Geiger-M\"uller counter. ``Messrs. M. Danysz and M. Zyw, working in this laboratory, have bombarded diverse substances with $\alpha$-rays from a thin-walled glass tube containing some 15 millicuries of radon, and immediately afterwards have tested their activity with a Geiger M\"uller counter... As the effect was apparent only in nitrogen, we conclude that it consists in a transmutation of nitrogen of the Joliot type, the probable reactions being: $_7$N$^{14}$ + $_2\alpha^4$ = $_9$F$^{17}$ + neutron, $_9$F$^{17}$ = $_8$O$^{17}$ + positron.'' The measured half-life of 1.2 minutes is in reasonable agreement with the present value of 64.49(16)~s. It is interesting to note that Wertenstein is the sole author although M. Danysz and M. Zyw performed the actual experiments.

\subsubsection*{$^{18}$F}

DuBridge et al. described the observation of $^{18}$F in 1937 in ``Proton Induced Radioactivity in Oxygen'' \cite{1937DuB01}. Protons up to 3.8 MeV impinged on quartz and other solid oxide targets and the resulting activities were measured following chemical separation. ``The second period of 107$\pm$4 min. is shown by chemical separation to be due also to an isotope of fluorine and is close to the 112 min. period found by Snell for $^{18}$F. This period must be attributed to the reaction O$^{18}$ + H$^1 \to$ F$^{18}$ + n$^1$, F$^{18} \to$ O$^{18}$ + e$^+$.'' This half-life agrees with the currently adopted value of 109.771(20)~min. The publication by Snell mentioned in the quote refers to a conference abstract \cite{1937Sne02}.

\subsubsection*{$^{19}$F}

Aston discovered $^{19}$F in 1920 as reported in ``The constitution of the elements'' \cite{1920Ast02}. The isotopes were identified by measuring their mass spectra. ``Fluorine (atomic weight 19.00) is apparently simple, as its chemical atomic weight would lead one to expect.''

\subsubsection*{$^{20}$F}

Crane et al. discovered $^{20}$F in 1935 as reported in the paper ``The Emission of Negative Electrons from Lithium and Fluorine Bombarded with Deuterons'' \cite{1935Cra02}. Deuterons of 0.8 MeV bombarded a fluorine target inside a cloud chamber and $^{20}$F was probably formed in the reaction $^{19}$F(d,p). Photographs of the electron tracks were taken and the energy distribution extracted. ``The electrons probably arise from the reaction F$^{19}$ + H$^2 \to$ F$^{20}$ + H$^1 \to$ Ne$^{20}$ + e$^-$ + H$^1$. We have measured the half-life of the radio-fluorine by means of an ionization chamber and found it to be 12$\pm$2 seconds.'' This half-life agrees with the currently accepted value of 11.163(8)~s.

\subsubsection*{$^{21}$F}

$^{21}$F was discovered in 1955 by Jarmie in ``Mass Measurement and Excited States of F$^{21}$'' \cite{1955Jar01}. CaF$_2$ and PbF$_2$ targets were bombarded with 1.82 MeV tritons from one of the Los Alamos 2.5 MeV Van de Graaff accelerators and $^{21}$F was identified by analyzing the proton ejectiles with a Cal-Tech type 16-inch double-focusing magnetic spectrometer. ``We wish to report that we have experimentally determined that F$^{21}$ is heavy-particle-stable and has a mass defect of 6.125$\pm$0.030 Mev or an atomic mass of 21.005703$\pm$0.000025 amu.''

\subsubsection*{$^{22}$F}

$^{22}$F was observed by Vaughn et al. in 1965 in ``New Isotope of Fluorine: F$^{22}$'' \cite{1965Vau01}. Neutrons of 14.8 MeV from the reaction T(d,n)$^4$He produced by the Lockheed 3.5 MeV Van de Graaff accelerator irradiated a $^{22}$Ne gas sample. $^{22}$F was formed in the (n,p) charge exchange reaction. Gamma- and $\beta$-rays of the activated samples were measured with NaI(Tl) scintillators and a plastic scintillator $\Delta$E-E telescope, respectively. ``Measurements of the half-life of F$^{22}$ were made by observing the number of $\beta$ rays emitted from F$^{22}$ (with and without a $\gamma$-ray coincidence requirement) as a function of time. The observed half-life is 4.0$\pm$0.4 seconds.'' This half-life agrees with the presently accepted value of 4.23(4)~s.

\subsubsection*{$^{23,24}$F}

Artukh et al. reported the first identification of $^{23}$F and $^{24}$F in 1970 in``New isotopes $^{23}$F, $^{24}$F, $^{25}$Ne and $^{26}$Ne, produced in nuclear reactions with heavy ions'' \cite{1970Art01}. A metallic $^{232}$Th target was bombarded with a 174 MeV $^{22}$Ne beam from the 300 cm heavy-ion cyclotron at Dubna, Russia. The reaction products were identified in a $\Delta$E-E semiconductor telescope at the focal plane of a magnetic spectrometer. ``The results quoted above show that apart from a number of known isotopes four new isotopes: $^{23}$F (720 events), $^{24}$F (25 events), $^{25}$Ne (400 events) and $^{26}$Ne (26 events) were obtained and identified unambiguously.''

\subsubsection*{$^{25}$F}

In 1970 $^{25}$F was discovered by Artukh et al. in ``New isotopes $^{21}$N, $^{23}$O, $^{24}$O and $^{25}$F, produced in nuclear reactions with heavy ions'' \cite{1970Art02}. A metallic $^{232}$Th target was bombarded with a 174 MeV $^{22}$Ne beam from the 310 cm heavy ion cyclotron at Dubna, Russia. The reaction products were identified in a $\Delta$E-E semiconductor telescope at the focal plane of a magnetic spectrometer. ``[The figure] shows that apart from a number of already known isotopes, four new isotopes: $^{21}$N (about 60 events), $^{23}$O (about 130 events), $^{24}$O (about 30 events) and $^{25}$F (about 40 events) have been obtained.''

\subsubsection*{$^{26}$F}

The first observation of $^{26}$F was reported by Westfall et al. in ``Production of neutron-rich nuclides by fragmentation of 212-MeV/amu $^{48}$Ca'' in 1979 \cite{1979Wes01}. $^{48}$Ca ions (212 MeV/nucleon) from the Berkeley Bevalac were fragmented on a beryllium target. The fragments were selected by a zero degree spectrometer and identified in a telescope consisting of 12 Si(Li) detectors, 2 position-sensitive Si(Li) detectors, and a veto scintillator. ``There is clear evidence for the particle stability of $^{22}$N, $^{26}$F, $^{33,34}$Al, $^{37,38,39}$Si, $^{40,41,42}$P, $^{41,42,43,44}$Si, and $^{44,45}$Cl with more than ten counts in each case.''

\subsubsection*{$^{27}$F}

$^{27}$F was discovered by Stevenson and Price in the 1981 paper ``Production of the neutron-rich nuclides $^{20}$C and $^{27}$F by fragmentation of 213 MeV/nucleon $^{48}$Ca'' \cite{1981Ste01}. $^{48}$Ca at 213 MeV/nucleon from the Berkeley Bevatron was fragmented on a beryllium target. The fragments were focussed on a stack of Lexan plastic track detectors in the zero-degree magnetic spectrometer. ``There is clear evidence for the first observation of $^{20}$C ($\sim$40 counts) and $^{27}$F ($\sim$20 counts).''

\subsubsection*{$^{29}$F}

Guillemaud-Mueller et al. announced the discovery of $^{29}$F in the 1989 article ``Observation of new neutron rich nuclei'' \cite{1989Gui01}. A 55~MeV/u $^{48}$Ca beam was fragmented on a tantalum target at GANIL and the projectile-like fragments were separated by the zero degree doubly achromatic LISE spetrometer. ``[The figure] shows  a part of a two-dimensional $\delta$E versus time-of-flight (i.e. A/Z) representation after a 15 h run with an average beam intensity of 200 enA. The known nitrogen isotopes $^{20}$N, $^{21}$N, $^{22}$N, $^{23}$N are clearly seen as well as $^{23}$O, $^{24}$O and the new isotope $^{29}$F (4 counts only).''

\subsubsection*{$^{31}$F}

$^{31}$F was discovered  by Sakurai et al. in 1999 as reported in ``Evidence for particle stability of $^{31}$F and particle instability of $^{25}$N and $^{28}$O'' \cite{1999Sak01}. A $^{40}$Ar beam was accelerated at the AVF and RIKEN Ring Cyclotron to 94.1 MeV/nucleon and was fragmented on a tantalum target. The fragments were analyzed by the RIPS spectrometer and identified on the basis of energy loss, total kinetic energy, time-of-flight and magnetic rigidity. ``The number of events obtained was 4387, 7072 and 905 for the isotopes $^{19}$B, $^{23}$N and $^{22}$C, respectively. A new isotope $^{31}$F (8 events) was observed for the first time.''

\subsection{Neon}\vspace{0.0cm}

The observation of 18 neon isotopes has been reported so far, including 3 stable, 3 proton-rich, 11 neutron-rich, and 1 proton-unbound resonance. The one-neutron unbound resonances of $^{33}$Ne and $^{35}$Ne should be able to be observed in the future. In addition, $^{36}$Ne still might be particle-stable.

\subsubsection*{$^{16}$Ne}

Holt et al. reported the first observation of $^{16}$Ne in the 1977 paper ``Pion non-analog double charge exchange: $^{16}$O($\pi^+,\pi^-$)$^{16}$Ne'' \cite{1977Hol01}. Pions from the Los Alamos Meson Physics Facility LAMPF bombarded a natural water target having the form of a thick gelatin disk. $^{16}$Ne was produced in the double-charge exchange reaction ($\pi^+,\pi^-$). The negative pions were analyzed with a zero degree spectrometer and identified by measuring position, velocity, energy-loss, and time-of-flight. ``The spectrum obtained at 0$^\circ$ for the $^{16}$O($\pi^+,\pi^-$)$^{16}$Ne reaction with incident pion of 145 MeV mean kinetic energy is shown [in the figure]. A distinct peak is observed near the energy predicted on the basis of mass systematics... These are expected inasmuch as $^{16}$Ne is approximately 2 MeV unbound with respect to two-proton decay to $^{14}$O.''

\subsubsection*{$^{17}$Ne}

In ``Observation of delayed proton radioactivity'' Barton et al. implied the observation of $^{17}$Ne for the first time in 1963 \cite{1963Bar01}. The McGill Synchrocyclotron accelerated protons to 97 MeV which bombarded LiF and NaF targets. $^{17}$Ne was identified by the observation of $\beta$-delayed protons in a silicon junction particle detector. ``The hypothesis that the decay of the nuclide (2k+2,2k-1) will dominate the delayed proton spectrum from targets of both element (2k+2) and element (2k+1) seems to be verified. In particular, by following very reasonable rules for predicting proton lines, all the observed lines are accounted for and all those predictions based on known level properties are borne out. The existence of Mg$^{21}$, Ne$^{17}$, and O$^{13}$ is assumed since it seems fairly certain that proton lines in the decay of each have been observed.'' The half-life was subsequently independently measured by D'Auria and Preiss \cite{1964DAu01} and McPherson et al. \cite{1964McP01}. Both papers acknowledged the first observation of $\beta$-delayed protons from $^{17}$Ne by Barton et al..

\subsubsection*{$^{18}$Ne}

$^{18}$Ne was discovered by Gow and Alvarez in ``Neon-18'' in 1954 \cite{1954Gow01}. Protons from the Berkeley linear accelerator bombarded teflon (CF$_2$)$_N$ and LiF crystal targets. $^{18}$Ne was produced in the reaction $^{19}$F(p,2n) and identified with a 180$^\circ$ magnetic spectrograph with two proportional counters in coincidence. ``A decay curve taken at the magnet current  that gave the best ratio of 1.6-sec activity to background is shown in [the figure]. From this we assign a value of the half-life of 1.6$\pm$0.2 sec.`` This half-life agrees with the presently adapted value of 1.672(8)~s.

\subsubsection*{$^{19}$Ne}

The observation of $^{19}$Ne was first reported by Fox et al. in 1939 in ``The difference in Coulomb Energy of Light, Isobaric Nuclei'' \cite{1939Fox01}. At Princeton, protons bombarded a fluorine target and $^{19}$Ne was formed in the (p,n) charge exchange reaction. The decay and absorption curves of the positrons were measured. ``We have commenced a systematic study of these nuclei and have found that the reaction F$^{19}$(p,n)Ne$^{19}$ gives a radioactivity of approximately the expected half-life and upper limit. The measured half-life is 20 sec. and the absorption curve of the positrons indicates an upper limit of 2.5 Mev.'' This half-life is close to the present value of 17.296(5)~s.

\subsubsection*{$^{20}$Ne}

The 1913 paper ``Bakerian Lecture: - Rays of Positive Electricity'' by Thomson can be considered the discovery of $^{20}$Ne as an isotope \cite{1913Tho01}. It represented the first observation of isotopes in a stable element. An electric discharge tube ionized particles of low pressure gases. Ions leaving through an aperture then pass through an electric and magnetic field and were deflected depending on their mass and velocity. ``The photograph shows that, in addition to helium and neon, there is another gas with an atomic weight about 22. This gas has been found in every specimen of neon which has been examined, including a very carefully purified sample prepared by Mr. E.W. Watson and a specimen very kindly supplied by M. Claud, of Paris... The substance giving the line 22 also occurs with a double charge, giving a line for which m/e = 11. There can, therefore, I think, be little doubt that what has been called neon is not a simple gas but a mixture of two gases, one of which has an atomic weight about 20 and the other about 22''.

\subsubsection*{$^{21}$Ne}

Hogness and Kvalnes reported the first observation of $^{21}$Ne in the 1928 paper ``The ionization processes in methane interpreted by the mass spectrograph'' \cite{1928Hog01}. In the process of studying CH$_4$ molecules with a mass spectrograph, neon was also analyzed. ``The calibrating gas, neon, gave the positive ions, Ne$_{20}^+$ and Ne$_{22}^+$ in the ratio of ten to one as found by Aston and by Barton and Bartlett. There was, in addition to these ions, a small peak for Ne$_{21}^+$, which was always found when neon was in the apparatus but was never obtained in the absence of neon.'' Aston had observed an ``extremely faint'' line at mass 21 but did not claim the existence of $^{21}$Ne \cite{1920Ast04}.

\subsubsection*{$^{22}$Ne}

$^{22}$Ne was first identified by Thomson in ``Bakerian Lecture: - Rays of Positive Electricity'' \cite{1913Tho01} in 1913. An electric discharge tube ionized particles of low pressure gases. Ions leaving through an aperture then pass through an electric and magnetic field and were deflected depending on their mass and velocity. ``The photograph shows that, in addition to helium and neon, there is another gas with an atomic weight about 22. This gas has been found in every specimen of neon which has been examined, including a very carefully purified sample prepared by Mr. E.W. Watson and a specimen very kindly supplied by M. Claud, of Paris... The substance giving the line 22 also occurs with a double charge, giving a line for which m/e = 11. There can, therefore, I think, be little doubt that what has been called neon is not a simple gas but a mixture of two gases, one of which has an atomic weight about 20 and the other about 22''. Thomson had reported the observation of the mass 22 line already a few month earlier, however, at that time he did not yet draw the conclusion that the line belongs to neon: ``The origin of this line presents many points of interest; there are no known gaseous compounds of any of the recognised elements which have this molecular weight. Again, if we accept Mendel\'eef's periodic law, there is no room for a new element with this atomic weight. There is, however, the possibility that we may be interpreting Mendel\'eef's law too rigidly, and that in the neighbourhood of the atomic weight of neon there may be a group of two or more elements with similar properties, just as in another part of the table we have the group iron, nickel, and cobalt. From the relative intensities of the 22 line and the neon line we may conclude that the quantity of the gas giving the 22 line is only a small fraction of the quantity of neon.''  \cite{1913Tho02}.

\subsubsection*{$^{23}$Ne}

In 1936 Nahmias and Walen identified $^{23}$Ne for the first time in ``Sur quelques radio\'el\'ements artificiels'' \cite{1936Nah01}. Neutrons from a Rn-Be source irradiated a laminated sodium metal sample and activities of 8(1)~s and 33(1)~s were observed.``Pour le Na la petite p\'eriode serait due \`a la r\'eaction $^{23}$Na + $^1$n $\to ^{20}$F + $^4$He et celle de 33 secondes \`a $^{23}$Na + $^1$n $\to ^{23}$Ne + $^1$H.'' [For the sodium, the short period would be due to the reaction  $^{23}$Na + $^1$n $\to ^{20}$F + $^4$He and the one of 33 seconds due to $^{23}$Na + $^1$n $\to ^{23}$Ne + $^1$H.] The half-life is close to the currently adopted value of 26.464(9)~s.

\subsubsection*{$^{24}$Ne}

In the 1956 article ``Decay of the new nuclide Ne$^{24}$'' Dropesky and Schardt reported the discovery of $^{24}$Ne \cite{1956Dro01}. A neon gas target was bombarded with 1.5 MeV tritons from the 2.5 MeV Los Alamos electrostatic accelerator and $^{24}$Ne was formed in the $^{22}$Ne(t,p) reaction. Beta- and $\gamma$-rays spectrometers recorded the activities after chemical separation. ``Treatment of the decay data by least squares analysis gave a value of 3.38$\pm$0.02 min for the half-life of Ne$^{24}$.'' This half-life still corresponds to the currently accepted value.

\subsubsection*{$^{25,26}$Ne}

Artukh et al. reported the first identification of $^{25}$Ne and $^{26}$Ne in 1970 in``New isotopes $^{23}$F, $^{24}$F, $^{25}$Ne and $^{26}$Ne, produced in nuclear reactions with heavy ions'' \cite{1970Art01}. A metallic $^{232}$Th target was bombarded with a 174 MeV $^{22}$Ne beam from the 300 cm heavy-ion cyclotron at Dubna, Russia. The reaction products were identified in a $\Delta$E-E semiconductor telescope at the focal plane of a magnetic spectrometer. ``The results quoted above show that apart from a number of known isotopes four new isotopes: $^{23}$F (720 events), $^{24}$F (25 events), $^{25}$Ne (400 events) and $^{26}$Ne (26 events) were obtained and identified unambiguously.''

\subsubsection*{$^{27}$Ne}

$^{27}$Ne was discovered by Butler et al. in ``Observation of the new nuclides $^{27}$Ne, $^{31}$Mg, $^{32}$Mg, $^{34}$Al, and $^{39}$P'' in 1977 \cite{1977But01}. $^{27}$Ne was produced in the spallation reaction of 800 MeV protons from the Clinton P. Anderson Meson Physics Facility LAMPF on a uranium target. The spallation fragments were identified with a silicon $\Delta$E-E telescope and by time-of-flight measurements. ``All of the stable and known neutron-rich nuclides (except $^{24}$O and the more neutron-rich Na isotopes) are seen. The five previously unobserved neutron-rich nuclides $^{27}$Ne, $^{31}$Mg, $^{32}$Mg, $^{34}$Al, and $^{39}$P are clearly evident. Each of these peaks contains ten or more events.''

\subsubsection*{$^{28}$Ne}

In 1979 Symons et al. described the discovery of $^{28}$Ne in ``Observation of new neutron-rich isotopes by fragmentation of 205-MeV/Nucleon $^{40}$Ar ions'' \cite{1979Sym01}. A 205 MeV/nucleon $^{40}$Ar beam from the Berkeley Bevalac was fragmented on a carbon target. The projectile fragments were analyzed with a zero-degree magnetic spectrometer and detected in two detector telescopes. ``Projected mass spectra with a gate of $\pm$0.2 units about charges 10, 11, 12, and 13 are shown in [the figure]. $^{28}$Ne and $^{35}$Al are positively identified as particle-stable isotopes with more than 10 counts in each case.''

\subsubsection*{$^{29,30}$Ne}

The first observation of $^{29}$Ne and $^{30}$Ne was reported in 1985 by Langevin et al. in ``Production of neutron-rich nuclei at the limits of particles stability by fragmentation of 44 MeV/u $^{40}$Ar projectiles'' \cite{1985Lan01}. A 44 MeV/u $^{40}$Ar beam was fragmented on a tantalum target at GANIL. The fragments were measured with the triple-focusing magnetic spectrometer LISE and identified by measuring energy-loss, energy and time-of-flight. ``The first observation of $^{23}$N, $^{29}$Ne and $^{30}$Ne and their particle bound character results clearly from the mass histograms of [the figures].''

\subsubsection*{$^{31}$Ne}

$^{31}$Ne was discovered by Sakurai et al. in 1996 as reported in ``Production and identification of new neutron-rich nuclei, $^{31}$Ne and $^{37}$Mg, in the reaction 80A MeV $^{50}$Ti + $^{181}$Ta'' \cite{1996Sak01}. A $^{50}$Ti beam was accelerated at the RIKEN Ring Cyclotron to 80 MeV/nucleon and fragmented on a tantalum target. The fragments were analyzed by the RIPS spectrometer and identified on the basis of energy loss, total kinetic energy, time-of-flight and magnetic rigidity. ``All of the fragments of $^{30,31,32}$Ne, $^{32,33,34,35}$Na, and $^{35,36,37}$Mg were stopped at the SSD4 with the selected window of the magnetic rigidity. Significant numbers of events have been observed for new isotopes, $^{31}$Ne (23 events) and $^{37}$Mg (three events).'' $^{31}$Ne had previously been reported incorrectly as particle unstable \cite{1990Gui01}.

\subsubsection*{$^{32}$Ne}

Guillemaud-Mueller et al. announced the discovery of $^{32}$Ne in the 1990 article ``Particle stability of the isotopes $^{26}$O and $^{32}$Ne in the reaction 44 MeV/nucleon $^{48}$Ca + Ta'' \cite{1990Gui01}. A 44~MeV/u $^{48}$Ca beam was fragmented on a tantalum target at GANIL and the projectile-like fragments were separated by the zero degree triple-focusing magnetic analyzer LISE. ``[The figure] represents the two-dimensional plot (Z versus time of flight) obtained under these conditions after a 40-h measurement with an average beam intensity of 160 enA. The heaviest known isotopes $^{19}$B, $^{22}$C, $^{29}$F, and the previously unknown isotope $^{32}$Ne (four events) are clearly visible.''

\subsubsection*{$^{34}$Ne}

In the 2002 article ``New neutron-rich isotopes, $^{34}$Ne, $^{37}$Na and $^{43}$Si, produced by fragmentation of a 64A MeV $^{48}$Ca beam'' Notani et al. described the first observation of $^{34}$Ne \cite{2002Not01}. The RIKEN ring cyclotron accelerated a $^{48}$Ca beam to 64 MeV/nucleon which was then fragmented on a tantalum target. The projectile fragments were analyzed with the RIPS spectrometer. ``[Part (a) of the figure] shows a two-dimensional plot of A/Z versus Z, obtained from the data accumulated with the $^{40}$Mg B$\rho$ setting, while [part (b)] is for the $^{43}$Si setting. The integrated beam intensities for the two settings are 6.9$\times$10$^{16}$ and 1.7$\times$10$^{15}$ particles, respectively. The numbers of events observed for three new isotopes, $^{34}$Ne, $^{37}$Na and $^{43}$Si, were 2, 3 and 4, respectively.'' Lukyanov et al. reported the discovery of $^{34}$Ne independently less than two months later \cite{2002Luk01}.

\section{Summary}
The discoveries of the known isotopes and unbound resonances have been compiled and the methods of their production discussed. 125 isotopes were described including 20 stable, 14 proton-rich, 61 neutron-rich, 14 proton- and 15 neutron-unbound resonances, plus $^8$Be breaking up into two $\alpha$-particles.

Historically, the light mass region is especially interesting. In 1908 Rutherford characterized the $\alpha$-particle correctly as a $^4$He nucleus. The existences of isotopes was first realized by Thomson with the observation of the two neon isotopes $^{20}$Ne and $^{22}$Ne in 1913. Aston used a mass spectrograph to determine isotopic mass identification for the first time in 1919 and reported $^{12}$C and $^{16}$O. $^{17}$O was the first new isotope produced in a nuclear reaction in 1925. The first isotope produced with an accelerator was $^8$Be in 1932 and finally the first observed artificial radioactivity is credited to Curie with the observation of $^{13}$N in 1933.

The discovery of the particle-stable isotopes was fairly uncontroversial, only the half-lives of $^8$He, $^{12}$Be, and $^{10}$C were initially measured incorrectly. $^{14}$Be and $^{31}$Ne were first reported to be unstable with respect to neutron emission. The neutron-unbound nuclei $^5$H, $^{21}$C, and $^{25}$O were initially reported to be bound. The assignment of discoveries for several of neutron-unbound nuclei can only be considered tentative. Further experiments are certainly needed to confirm the presence of the various resonances.

\ack

I would like to thank John Kelley for verifying the assignments of the discoveries and John Kelley and Ute Thoennessen for carefully proofreading the manuscript. This work was supported by the National Science Foundation under grant No. PHY06-06007 (NSCL).

\bibliography{../isotope-discovery-references}

\newpage

\newpage

\TableExplanation

\bigskip
\renewcommand{\arraystretch}{1.0}

\section*{Table 1.\label{tbl1te} Discovery of light isotopes with Z $\le$ 10}
\begin{tabular*}{0.95\textwidth}{@{}@{\extracolsep{\fill}}lp{5.5in}@{}}
\multicolumn{2}{p{0.95\textwidth}}{  }\\

Isotope &  Name of isotope \\
First Author & First author of refereed publication \\
Journal & Journal of publication \\
Ref. &  Reference  \\
Method & Production method used in the discovery: \\
    & AS: atomic spectroscopy \\
    & CR: cathode rays \\
    & MS: mass spectroscopy \\
    & DI: deep inelastic reactions \\
    & FE: fusion evaporation \\
    & LP: light-particle reactions (including neutrons) \\
    & PF: projectile fragmentation \\
    & PI: pion-induced reactions \\
    & SB: reactions with secondary beams \\
    & SP: spallation reactions \\
    & TR: heavy-ion transfer reactions \\
Laboratory &  Laboratory where the experiment was performed\\
Country &  Country of laboratory\\
Year & Year of discovery  \\
\end{tabular*}
\label{tableI}

\datatables 



\setlength{\LTleft}{0pt}
\setlength{\LTright}{0pt}


\setlength{\tabcolsep}{0.5\tabcolsep}

\renewcommand{\arraystretch}{1.0}

\footnotesize 

\begin{longtable}{@{\extracolsep\fill}rllrllll@{}}
\caption{Discovery of Light Isotopes with Z $\le$ 10. See page\ \pageref{tbl1te} for explanation of table}
Isotope & First Author & Journal & Ref. & Method & Laboratory & Country & Year\\
\hline\\
\endfirsthead\\
\caption[]{(continued)}
Isotope & Author & Journal & Ref. & Method & Laboratory & Country & Year\\
\hline\\
\endhead
$^1$n& J. Chadwick & Nature &\cite{1932Cha01}& LP & Cambridge & UK &1932 \\
$^2$n& R.P. Haddock & Phys. Rev. Lett. &\cite{1965Had01}& PI & Berkeley & USA &1965 \\
        &            &               &                &    &       &       &     \\
        &            &               &                &    &       &       &     \\
$^1$H& F.W. Aston & Nature &\cite{1920Ast01}& MS & Cambridge & UK &1920 \\
$^2$H& H.C. Urey & Phys. Rev. &\cite{1932Ure01}& AS & Columbia & USA &1932 \\
$^3$H& M.L. Oliphant & Nature &\cite{1934Oli01}& LP & Cambridge & UK &1934 \\
$^4$H& U. Sennhauser & Phys. Lett. B &\cite{1981Sen01}& PI & Zurich & Switzerland &1981 \\
$^5$H& M.G. Gornov & JETP Lett. &\cite{1987Gor01}& PI & Leningrad & Russia &1987 \\
$^6$H& D.V. Aleksandrov & Sov. J. Nucl. Phys. &\cite{1984Ale01}& TR & Moscow & Russia &1984 \\
$^7$H& A.A. Korsheninnikov & Phys. Rev. Lett. &\cite{2003Kor01}& SB & RIKEN & Japan &2003 \\
        &            &               &                &    &       &       &     \\
        &            &               &                &    &       &       &     \\
$^2$He& M.A. Tuve & Phys. Rev. &\cite{1936Tuv01}& LP & Carnegie Institute & USA &1936 \\
$^3$He& M.L. Oliphant & Nature &\cite{1934Oli01}& LP & Cambridge & UK &1934 \\
$^4$He& E. Rutherford & Proc. Roy. Soc. &\cite{1908Rut01}& MS & Manchester & UK &1908 \\
$^5$He& J.H. Williams & Phys. Rev. &\cite{1937Wil01}& LP & Minnesota & USA &1937 \\
$^6$He& T. Bjerge & Nature &\cite{1936Bje01}& LP & Copenhagen & Denmark &1936 \\
$^7$He& R.H. Stokes & Phys. Rev. Lett. &\cite{1967Sto01}& LP & Los Alamos & USA &1967 \\
$^8$He& A.M. Poskanzer & Phys. Rev. Lett. &\cite{1965Pos01}& SP & Brookhaven & USA &1965 \\
$^9$He& K.K. Seth & Phys. Rev. Lett. &\cite{1987Set01}& PI & Los Alamos & USA &1987 \\
$^{10}$He& A.A. Korsheninnikov & Phys. Lett. B &\cite{1994Kor01}& SB & RIKEN & Japan &1994 \\
        &            &               &                &    &       &       &     \\
        &            &               &                &    &       &       &     \\
$^4$Li& J. Cerny & Phys. Rev. Lett. &\cite{1965Cer01}& LP & Berkeley & USA &1965 \\
$^5$Li& N.P. Heydenburg & Phys. Rev. &\cite{1941Hey01}& LP & Carnegie Institute & USA &1941 \\
$^6$Li& F.W. Aston & Nature &\cite{1921Ast01}& MS & Cambridge & UK &1921 \\
$^7$Li& F.W. Aston & Nature &\cite{1921Ast01}& MS & Cambridge & UK &1921 \\
$^8$Li& H.R. Crane & Phys. Rev. &\cite{1935Cra02}& LP & Caltech& USA &1935 \\
$^9$Li& W.L. Gardner & Phys. Rev. &\cite{1951Gar01}& LP & Berkeley & USA &1951 \\
$^{10}$Li& K.H. Wilcox & Phys. Lett. B &\cite{1975Wil01}& TR & Berkeley & USA &1975 \\
$^{11}$Li& A.M. Poskanzer & Phys. Rev. Lett. &\cite{1966Pos01}& SP & Berkeley & USA &1966 \\
$^{12}$Li& Yu. Aksyutina & Phys. Lett. B &\cite{2008Aks01}& SB & Darmstadt & Germany &2008 \\
$^{13}$Li& Yu. Aksyutina & Phys. Lett. B &\cite{2008Aks01}& SB & Darmstadt & Germany &2008 \\
        &            &               &                &    &       &       &     \\
        &            &               &                &    &       &       &     \\
$^6$Be& G.F. Bogdanov & J. Nucl. Ener. &\cite{1958Bog01}& LP & Moscow & Russia &1958 \\
$^7$Be& R.B. Roberts & Phys. Rev. &\cite{1938Rob01}& LP & Carnegie Institute & USA &1938 \\
$^8$Be& J.D. Cockcroft & Nature &\cite{1932Coc01}& LP & Cambridge & UK &1932 \\
$^9$Be& G.P. Thomson & Nature &\cite{1921Tho01}& MS & Cambridge & UK &1921 \\
$^{10}$Be& M.L. Oliphant & Proc. Roy. Soc. &\cite{1935Oli01}& LP & Cambridge & UK &1935 \\
$^{11}$Be& M.J. Nurmia & Phys. Rev. Lett. &\cite{1958Nur01}& LP & Arkansas & USA &1958 \\
$^{12}$Be& A.M. Poskanzer & Phys. Rev. Lett. &\cite{1966Pos01}& SP & Berkeley & USA &1966 \\
$^{13}$Be& D.V. Aleksandrov & Sov. J. Nucl. Phys. &\cite{1983Ale01}& TR & Moscow & Russia &1983 \\
$^{14}$Be& J.D. Bowman & Phys. Rev. Lett. &\cite{1973Bow01}& SP & Berkeley & USA &1973 \\
        &            &               &                &    &       &       &     \\
        &            &               &                &    &       &       &     \\
$^7$B& R.L. McGrath & Phys. Rev. Lett. &\cite{1967McG01}& LP & Berkeley & USA &1967 \\
$^8$B& L.W. Alvarez & Phys. Rev. &\cite{1950Alv01}& LP & Berkeley & USA &1950 \\
$^9$B& R.O. Haxby & Phys. Rev. &\cite{1940Hax01}& LP & Westinghouse & USA &1940 \\
$^{10}$B& F.W. Aston & Nature &\cite{1920Ast02}& MS & Cambridge & UK &1920 \\
$^{11}$B& F.W. Aston & Nature &\cite{1920Ast02}& MS & Cambridge & UK &1920 \\
$^{12}$B& H.R. Crane & Phys. Rev. &\cite{1935Cra01}& LP & Caltech& USA &1935 \\
$^{13}$B& S.K. Allison & Phys. Rev. &\cite{1956All01}& FE & Chicago & USA &1956 \\
$^{14}$B& A.M. Poskanzer & Phys. Rev. Lett. &\cite{1966Pos01}& SP & Berkeley & USA &1966 \\
$^{15}$B& A.M. Poskanzer & Phys. Rev. Lett. &\cite{1966Pos01}& SP & Berkeley & USA &1966 \\
$^{16}$B& R. Kalpakchieva & Eur. Phys. J. A &\cite{2000Kal01}& TR & Berlin & Germany &2000 \\
$^{17}$B& J.D. Bowman & Phys. Rev. Lett. &\cite{1973Bow01}& SP & Berkeley & USA &1973 \\
$^{18}$B& A. Spyrou & Phys. Lett. B &\cite{2010Spy01}& SB & Michigan State & USA &2010 \\
$^{19}$B& J.A. Musser & Phys. Rev. Lett. &\cite{1984Mus01}& PF & Berkeley & USA &1984 \\
        &            &               &                &    &       &       &     \\
        &            &               &                &    &       &       &     \\
$^8$C& R.G.H. Robertson & Phys. Rev. Lett. &\cite{1974Rob01}& LP & Juelich & Germany &1974 \\
$^9$C& J. Cerny & Phys. Rev. Lett. &\cite{1964Cer01}& LP & Berkeley & USA &1964 \\
$^{10}$C& R. Sherr & Phys. Rev. &\cite{1949She01}& LP & Princeton & USA &1949 \\
$^{11}$C& H.R. Crane & Phys. Rev. &\cite{1934Cra01}& LP & Caltech& USA &1934 \\
$^{12}$C& F.W. Aston & Nature &\cite{1919Ast01}& MS & Cambridge & USA &1919 \\
$^{13}$C& A.S. King & Phys. Rev. &\cite{1929Kin01}& AS & Carnegie Institute & USA &1929 \\
$^{14}$C& T.W. Bonner & Phys. Rev. &\cite{1936Bon01}& LP & Caltech& USA &1936 \\
$^{15}$C& E.L. Hudspeth & Phys. Rev. &\cite{1950Hud01}& LP & Carnegie Institute & USA &1950 \\
$^{16}$C& S. Hinds & Phys. Rev. Lett. &\cite{1961Hin01}& LP & Aldermaston & UK &1961 \\
$^{17}$C& A.M. Poskanzer & Phys. Lett. B &\cite{1968Pos01}& SP & Berkeley & USA &1968 \\
$^{18}$C& A.G. Artukh & Nucl. Phys. A &\cite{1969Art01}& TR & Dubna & Russia &1969 \\
$^{19}$C& J.D. Bowman & Phys. Rev. C &\cite{1974Bow01}& SP & Berkeley & USA &1974 \\
$^{20}$C& J.D. Stevenson & Phys. Rev. C &\cite{1981Ste01}& PF & Berkeley & USA &1981 \\
$^{21}$C&                &              &                &    &          &     &     \\
$^{22}$C& F. Pougheon & Europhys. Lett. &\cite{1986Pou01}& PF & GANIL & France &1986 \\
        &            &               &                &    &       &       &     \\
        &            &               &                &    &       &       &     \\
$^{10}$N& A. Lepine-Szily & Phys. Rev. C &\cite{2002Lep01}& TR & GANIL & France &2002 \\
$^{11}$N& W. Benenson & Phys. Rev. C &\cite{1974Ben01}& LP & Michigan State & USA &1974 \\
$^{12}$N& L.W. Alvarez & Phys. Rev. &\cite{1949Alv02}& LP & Berkeley & USA &1949 \\
$^{13}$N& I. Curie & Compt. Rend. Acad. Sci. &\cite{1934Cur01}& LP & Paris & France &1934 \\
$^{14}$N& F.W. Aston & Nature &\cite{1920Ast01}& MS & Cambridge & UK &1920 \\
$^{15}$N& S.M. Naude & Phys. Rev. &\cite{1929Nau01}& AS & Chicago & USA &1929 \\
$^{16}$N& W.D. Harkins & Phys. Rev. &\cite{1933Har01}& LP & Chicago & USA &1933 \\
$^{17}$N& L.W. Alvarez & Phys. Rev. &\cite{1949Alv01}& LP & Berkeley & USA &1949 \\
$^{18}$N& L.F. Chase & Phys. Rev. Lett. &\cite{1964Cha01}& LP & Lockeed Palo Alto & USA &1964 \\
$^{19}$N& T.D. Thomas & Phys. Lett. B &\cite{1968Tho01}& SP & Princeton & USA &1968 \\
$^{20}$N& A.G. Artukh & Nucl. Phys. A &\cite{1969Art01}& TR & Dubna & Russia &1969 \\
$^{21}$N& A.G. Artukh & Phys. Lett. B &\cite{1970Art02}& DI & Dubna & Russia &1970 \\
$^{22}$N& G.D. Westfall & Phys. Rev. Lett. &\cite{1979Wes01}& PF & Berkeley & USA &1979 \\
$^{23}$N& M. Langevin & Phys. Lett. B &\cite{1985Lan01}& PF & GANIL & France &1985 \\
        &            &               &                &    &       &       &     \\
        &            &               &                &    &       &       &     \\
$^{12}$O& G.J. KeKelis & Phys. Rev. C &\cite{1978KeK01}& LP & Berkeley & USA &1978 \\
$^{13}$O& R. Barton & Can. J. Phys. &\cite{1963Bar01}& LP & McGill & Canada &1963 \\
$^{14}$O& R. Sherr & Phys. Rev. &\cite{1949She01}& LP & Princeton & USA &1949 \\
$^{15}$O& M.S. Livingston & Phys. Rev. &\cite{1934Liv01}& LP & Berkeley & USA &1934 \\
$^{16}$O& F.W. Aston & Nature &\cite{1919Ast01}& MS & Cambridge & USA &1919 \\
$^{17}$O& P.M.S. Blackett & Proc. Roy. Soc. &\cite{1925Bla01}& LP & Cambridge & UK &1925 \\
$^{18}$O& W.F. Giauque & J. Am. Chem. Soc. &\cite{1929Gia01}& AS & Berkeley & USA &1929 \\
$^{19}$O& M.E. Nahmias & Compt. Rend. Acad. Sci. &\cite{1936Nah01}& LP & Paris & France &1936 \\
$^{20}$O& N. Jarmie & Phys. Rev. Lett. &\cite{1959Jar01}& LP & Los Alamos & USA &1959 \\
$^{21}$O& T.D. Thomas & Phys. Lett. B &\cite{1968Tho01}& SP & Princeton & USA &1968 \\
$^{22}$O& A.G. Artukh & Nucl. Phys. A &\cite{1969Art01}& TR & Dubna & Russia &1969 \\
$^{23}$O& A.G. Artukh & Phys. Lett. B &\cite{1970Art02}& DI & Dubna & Russia &1970 \\
$^{24}$O& A.G. Artukh & Phys. Lett. B &\cite{1970Art02}& DI & Dubna & Russia &1970 \\
$^{25}$O& C.R. Hoffman & Phys. Rev. Lett. &\cite{2008Hof01}& SB & Michigan State & USA &2008 \\
        &            &               &                &    &       &       &     \\
        &            &               &                &    &       &       &     \\
$^{14}$F& V.Z. Goldberg & Phys. Lett. B &\cite{2010Gol01}& SB & Texas A\&M & USA &2010 \\
$^{15}$F& G.J. KeKelis & Phys. Rev. C &\cite{1978KeK01}& LP & Berkeley & USA &1978 \\
$^{16}$F& H.C. Bryant & Nucl. Phys. &\cite{1964Bry01}& LP & Los Alamos & USA &1964 \\
$^{17}$F& L. Wertenstein & Nature &\cite{1934Wer01}& LP & Warsaw & Poland &1934 \\
$^{18}$F& L.A. DuBridge & Phys. Rev. &\cite{1937DuB01}& LP & Rochester & USA &1937 \\
$^{19}$F& F.W. Aston & Nature &\cite{1920Ast02}& MS & Cambridge & UK &1920 \\
$^{20}$F& H.R. Crane & Phys. Rev. &\cite{1935Cra02}& LP & Caltech& USA &1935 \\
$^{21}$F& N. Jarmie & Phys. Rev. &\cite{1955Jar01}& LP & Los Alamos & USA &1955 \\
$^{22}$F& F.J. Vaughn & Phys. Rev. Lett. &\cite{1965Vau01}& LP & Lockeed Palo Alto & USA &1965 \\
$^{23}$F& A.G. Artukh & Phys. Lett. B &\cite{1970Art01}& DI & Dubna & Russia &1970 \\
$^{24}$F& A.G. Artukh & Phys. Lett. B &\cite{1970Art01}& DI & Dubna & Russia &1970 \\
$^{25}$F& A.G. Artukh & Phys. Lett. B &\cite{1970Art02}& DI & Dubna & Russia &1970 \\
$^{26}$F& G.D. Westfall & Phys. Rev. Lett. &\cite{1979Wes01}& PF & Berkeley & USA &1979 \\
$^{27}$F& J.D. Stevenson & Phys. Rev. C &\cite{1981Ste01}& PF & Berkeley & USA &1981 \\
$^{28}$F&                &              &                &    &          &     &     \\
$^{29}$F& D. Guillemaud-Mueller & Z. Phys. A &\cite{1989Gui01}& PF & GANIL & France &1989 \\
$^{30}$F&                &              &                &    &          &     &     \\
$^{31}$F& H. Sakurai & Phys. Lett. B &\cite{1999Sak01}& PF & RIKEN & Japan &1999 \\
        &            &               &                &    &       &       &     \\
        &            &               &                &    &       &       &     \\
$^{16}$Ne& R.J. Holt & Phys. Lett. B &\cite{1977Hol01}& PI & Los Alamos & USA &1977 \\
$^{17}$Ne& R. Barton & Can. J. Phys. &\cite{1963Bar01}& LP & McGill & Canada &1963 \\
$^{18}$Ne& J.D. Gow & Phys. Rev. &\cite{1954Gow01}& LP & Berkeley & USA &1954 \\
$^{19}$Ne& J.G. Fox & Phys. Rev. &\cite{1939Fox01}& LP & Princeton & USA &1939 \\
$^{20}$Ne& J.J. Thomson & Nature &\cite{1913Tho01}& CR & Cambridge & UK &1913 \\
$^{21}$Ne& T.R. Hogness & Phys. Rev. &\cite{1928Hog01}& MS & Berkeley & USA &1928 \\
$^{22}$Ne& J.J. Thomson & Nature &\cite{1913Tho01}& CR & Cambridge & UK &1913 \\
$^{23}$Ne& M.E. Nahmias & Compt. Rend. Acad. Sci. &\cite{1936Nah01}& LP & Paris & France &1936 \\
$^{24}$Ne& B.J. Dropesky & Phys. Rev. &\cite{1956Dro01}& LP & Los Alamos & USA &1956 \\
$^{25}$Ne& A.G. Artukh & Phys. Lett. B &\cite{1970Art01}& DI & Dubna & Russia &1970 \\
$^{26}$Ne& A.G. Artukh & Phys. Lett. B &\cite{1970Art01}& DI & Dubna & Russia &1970 \\
$^{27}$Ne& G.W. Butler & Phys. Rev. Lett. &\cite{1977But01}& SP & Los Alamos & USA &1977 \\
$^{28}$Ne& T.J.M. Symons & Phys. Rev. Lett. &\cite{1979Sym01}& PF & Berkeley & USA &1979 \\
$^{29}$Ne& M. Langevin & Phys. Lett. B &\cite{1985Lan01}& PF & GANIL & France &1985 \\
$^{30}$Ne& M. Langevin & Phys. Lett. B &\cite{1985Lan01}& PF & GANIL & France &1985 \\
$^{31}$Ne& H. Sakurai & Phys. Rev. C &\cite{1996Sak01}& PF & RIKEN & Japan &1996 \\
$^{32}$Ne& D. Guillemaud-Mueller & Phys. Rev. C &\cite{1990Gui01}& PF & GANIL & France &1990 \\
$^{33}$Ne&                &              &                &    &          &     &     \\
$^{34}$Ne& M. Notani & Phys. Lett. B &\cite{2002Not01}& PF & RIKEN & Japan &2002 \\

\\
\end{longtable}

\end{document}